\pdfoutput=1

\documentclass[onecolumn]{emulateapj}

\usepackage{amsmath}
\usepackage{graphicx}
\usepackage{color}

\newcommand{\be}{\begin{eqnarray}}
\newcommand{\ee}{\end{eqnarray}}

\def\jcap{JCAP}

\shorttitle{Testing the Kerr black hole hypothesis}
\shortauthors{Bambi et al.}

\begin{document}

\title{Testing the Kerr black hole hypothesis using X-ray reflection spectroscopy}

\author{Cosimo~Bambi\altaffilmark{1,2}, Alejandro~C\'ardenas-Avenda\~no\altaffilmark{3}, Thomas~Dauser\altaffilmark{4}, Javier~A.~Garc\'ia\altaffilmark{5}, and Sourabh~Nampalliwar\altaffilmark{1}}

\altaffiltext{1}{Center for Field Theory and Particle Physics and Department of Physics, 
Fudan University, 200433 Shanghai, China}
\altaffiltext{2}{Theoretical Astrophysics, Eberhard-Karls Universit\"at T\"ubingen, 72076 T\"ubingen, Germany}
\altaffiltext{3}{Programa de Matem\'atica, Fundaci\'on Universitaria Konrad Lorenz, 110231 Bogot\'a, Colombia}
\altaffiltext{4}{Remeis Observatory \& ECAP, Universit\"at Erlangen-N\"urnberg, 96049 Bamberg, Germany}
\altaffiltext{5}{Harvard-Smithsonian Center for Astrophysics, Cambridge, MA 02138, United States}

%\altaffiltext{\dag}{Hubble Fellow}
%\email[Corresponding author:]{bambi@fudan.edu.cn}

\begin{abstract}
We present the first X-ray reflection model for testing the assumption that the metric of astrophysical black holes is described by the Kerr solution. We employ the formalism of the transfer function proposed by Cunningham. The calculations of the reflection spectrum of a thin accretion disk are split into two parts: the calculation of the transfer function and the calculation of the local spectrum at any emission point in the disk. The transfer function only depends on the background metric and takes into account all the relativistic effects (gravitational redshift, Doppler boosting, and light bending). Our code computes the transfer function for a spacetime described by the Johannsen metric and can easily be extended to any stationary, axisymmetric, and asymptotically flat spacetime. Transfer functions and single line shapes in the Kerr metric are compared with those calculated from existing codes to check that we reach the necessary accuracy. We also simulate some observations with NuSTAR and LAD/eXTP and fit the data with our new model to show the potential capabilities of current and future observations to constrain possible deviations from the Kerr metric.
\end{abstract}


%%%%%%%%%%%%%%%%%%%%%%%%%%%%%%%

\section{Introduction \label{s-1}}

The theory of general relativity was proposed by Einstein about a century ago and is still the standard framework for the description of the gravitational field and the chrono-geometrical structure of the spacetime. The first test of general relativity can be dated back to the measurement of light bending by the Sun by Eddington in 1919~\citep{edd1919}. Especially over the past 60~years, there have been significant efforts to test the theory in weak gravitational fields, mainly with precise experiments in the Solar System and accurate radio observations of binary pulsars~\citep{will}. Tests of general relativity in the strong gravity regime are nowadays the new frontier, both with electromagnetic radiation~\citep{er1,er2,j2016} and gravitational waves~\citep{gw1,gw2}.

Astrophysical black holes are the ideal laboratory for testing strong gravity. In 4-dimensional general relativity, an uncharged black hole is described by the Kerr solution\footnote{There are a number of assumptions behind this statement. In particular, the spacetime must have 4~dimensions and be stationary and asymptotically flat; the exterior must be regular (no singularities or closed time-like curves); the metric is a vacuum solution of the Einstein equations. See, e.g., \citet{nh-rev} for more details.} and is completely described by only two parameters, namely the mass $M$ and the spin angular momentum $J$ of the object. This is the result of the ``no-hair theorem''~\citep{nh1,nh2}. It is remarkable that the spacetime around astrophysical black holes should be well described by the Kerr metric. As soon as a black hole is formed, initial deviations from the Kerr solution are quickly radiated away with the emission of gravitational waves~\citep{k1}. The equilibrium electric charge is extremely small for macroscopic objects and completely negligible for the spacetime geometry~\citep{k2}. Accretion disks typically have a mass of several orders of magnitude smaller than the central object and their impact on the background metric can be safely ignored~\citep{k3,k4}.

Within Einstein's theory of gravity, the Kerr metric should well describe the spacetime around astrophysical black holes. Nevertheless, macroscopic deviations from the Kerr spacetime are possible in many scenarios. For instance, \citet{carlos} have recently discovered a family of hairy black holes in 4-dimensional Einstein gravity minimally coupled to a complex, massive scalar field. Hairy black holes generically arise when scalar fields non-minimally coupled to gravity, and an example is the dilaton in Einstein-dilaton-Gauss-Bonnet gravity~\citep{hbh-EdGB}. Quantum gravity effects might also produce macroscopic corrections to the Kerr metric~\citep{dvali1,dvali2,giddings}.

Electromagnetic and gravitational radiations can test general relativity in different ways. The properties of the electromagnetic radiation emitted by the accreting gas close to a black hole depend on both the gas motion in the strong gravity region and the photon propagation from the emission point in the disk to the detection point in the flat faraway region. In this case, we can test the Kerr metric as Solar System experiments have so far tested the Schwarzschild solution in the weak field limit. However, it is not possible to distinguish a Kerr black hole in general relativity from a Kerr black hole in an alternative metric theory of gravity, because the geodesic motion is the same~\citep{psaltis}. Gravitational waves can instead probe the field equations of the theory, while they are less suitable to perform model-independent tests. The two approaches can thus be seen as complementary; see, e.g., \citet{kz}, \citet{comp1}, and \citet{comp2}.

With the electromagnetic approach, currently there are two leading techniques to probe the strong gravity region around a black hole: the study of the thermal spectrum of thin disks (continuum-fitting method)~\citep{cfm1,cfm1b,cfm2} and the analysis of the relativistically smeared reflection spectrum of thin disks (reflection method)~\citep{iron1,iron1b,iron2}. Both techniques have been developed for measuring black hole spins under the assumption of Kerr background and can be naturally extended for testing the Kerr metric~\citep{cfm3,cfm4,cfm5,iron3,ss09,iron4,iron5,cfm-iron-c,iron6,yy16}.

The reflection method has a number of advantages with respect to the study of the thermal spectrum. It can be easily applied to both stellar-mass and supermassive black holes\footnote{The continuum-fitting method has also been applied to supermassive black holes, but only in very special cases~\citep[e.g.][]{czerny2011,done2013}.}. It is independent of the black hole mass and distance, while the inclination angle of the disk with respect to the line of sight of the observer can be inferred from the fit of the reflection spectrum; with the continuum-fitting method, these three quantities have to be obtained from other measurements and their uncertainty is often large. In the presence of high quality data and the correct astrophysical model, the reflection method is potentially quite a powerful tool to constrain the metric around black holes~\citep[see, for instance,][]{jjc1,jjc2,jjc3,comp1}.

Theoretical models of X-ray reflection have been undergoing active development over three decades~\citep[see][for a review]{fr2010}. Currently, the most advanced model is {\sc xillver}~\citep{xillver1,xillver2}, and its relativistic counterpart {\sc relxill}~\citep{relline2,relxill}. These are the state-of-the-art in modeling reflection in strong gravity.

Compared to all earlier reflection codes, {\sc xillver} provides a superior treatment of the radiative transfer, as well as an improved calculation of the ionization balance, by implementing the photoionization routines from the {\sc xstar} code~\citep{kallman2001}, which incorporates the most complete atomic database for modeling synthetic photoionized X-ray spectra. The microphysics captured by {\sc xillver} is much more rigorous than for any earlier code, principally because of the detailed treatment of the K-shell atomic properties of the prominent ions~\citep[e.g.,][]{garcia2005,kallman2004,garcia2009}.

The model {\sc relxill} is the result from the combination of {\sc xillver} with
the relativistic blurring code {\sc relconv}~\citep{relline1}. {\sc relconv} is a relativistic convolution code that, assuming the Kerr metric, requires as input the local spectrum at any emission point in the disk and gives as output the spectrum measured by a distant observer. The aim of our work here is to construct a model to extend {\sc relxill} to a generic stationary, axisymmetric and asymptotically flat black hole metric. We replace {\sc relconv} with a more general relativistic convolution code, while we maintain {\sc xillver} because the microphysics of the local spectrum does not change.

In this Paper, we present a new code to compute transfer functions in any stationary, axisymmetric, and asymptotically flat black hole metric and extend {\sc relxill} for testing the Kerr black hole hypothesis. Current studies along this line of research model the X-ray spectrum with a simple power-law plus a relativistically broadened iron line~\citep{jjc1,jjc2,jjc3}. This can be sufficient for a preliminary study and a qualitative analysis. However, this is definitively not adequate if we really want to test general relativity. Here we employ the formalism of the transfer function for thin accretion disks~\citep{cun75}. In this framework, the calculations of the reflection spectrum are split into two parts: the calculation of the transfer function and the calculation of the reflection spectrum in the rest-frame of the gas. The transfer function only depends on the metric of the background and takes into account all the relativistic effects (gravitational redshift, Doppler boosting, light bending). The local spectrum is obtained by solving radiation transfer on a plane-parallel, 1-dimensional slab and is not strictly related to the metric of the spacetime.

In order to test the Kerr metric, our model must be able to compute the X-ray reflection spectrum of a thin disk in a background more general than the Kerr solution and that includes the Kerr solution as a special case. The test-metric is described by the mass $M$ and the spin angular momentum $J$ of the object, as well as by a number of ``deformation parameters''. The latter are used to quantify possible deviations from the Kerr metric and are the parameters to constrain from observations to verify the Kerr black hole hypothesis. The Kerr metric is recovered when all the deformation parameters vanish, while there are deviations from the Kerr solution in the presence of at least one non-vanishing deformation parameter.

In the standard case of the Kerr metric, the calculations of the transfer function exploit some specific properties of the Kerr solution~\citep{cun75,spe95}. Because of the presence of the Carter constant, the equations of motions are separable. More importantly, the equations in the $(r,\theta)$ plane can be reduced to elliptic integrals. This significantly simplifies the calculations of the transfer function. In our more general case, the transfer function is evaluated by integrating the photon geodesic equations from the point of detection in the plane of the distant observer backward in time to the point of emission in the accretion disk. Our calculations are inevitably longer than those in the Kerr metric that solve elliptic integrals.

The Paper is organized as follows. In Section~\ref{s-2}, we review the formalism of the transfer function and, in Section~\ref{s-6}, the Johannsen metric~\citep{j-m}, which is the one adopted in our current version of the code. Section~\ref{s-4} describes our numerical method to compute the transfer function. In Section~\ref{s-5}, we compare transfer functions and single iron line shapes produced by our code for a few Kerr solutions with those calculated by existing codes. Section~\ref{s-5bis} shows some examples of transfer functions and single line shapes in the Johannsen metric. In Section~\ref{s-new}, we simulate several observations of a bright black hole binary with NuSTAR and LAD/eXTP and we fit the data with our new version of {\sc relxill} to constrain one of the deformation parameters in the Johannsen metric as an illustrative example of the application of the new model and the constraining power of current and future X-ray missions. Summary and conclusions are reported in Section~\ref{s-7}. In Appendix~\ref{s-3}, we present all the formulas to compute the transfer function for a thin accretion disk in a generic stationary, axisymmetric, and asymptotically flat black hole spacetime. Throughout the Paper, we employ units in which $G_{\rm N} = c = 1$ and the convention of a metric with signature $(-+++)$. Except in Section~\ref{s-6}, we set the black hole mass parameter $M$ as defined in the Kerr and Johannsen metrics equal to 1.

%%%%%%%%%%%%%%%%%%%%%%%%%%%%%%%

\section{Transfer function for thin accretion disks}\label{s-2}

In this section we review the formalism of the transfer function for geometrically thin and optically thick accretion disks~\citep{cun75,spe95}. The observed flux of a thin accretion disk (measured, for instance, in erg~s$^{-1}$~cm$^{-2}$~Hz$^{-1}$) can be written as
\be\label{eq-thin-Fobs}
F_{\rm o} (\nu_{\rm o}) = \int I_{\rm o}(\nu_{\rm o}, X, Y) d\tilde{\Omega} 
= \int g^3 I_{\rm e}(\nu_{\rm e}, r_{\rm e}, \vartheta_{\rm e}) d\tilde{\Omega} \, .
\ee
$I_{\rm o}$ and $I_{\rm e}$ are, respectively, the specific intensity of the radiation detected by the distant observer and the specific intensity of the radiation as measured by the emitter (for instance, in erg~s$^{-1}$~cm$^{-2}$~str$^{-1}$~Hz$^{-1}$). $X$ and $Y$ are the Cartesian coordinates of the image of the disk in the plane of the distant observer. $d\tilde{\Omega} = dX dY/D^2$ is the element of the solid angle subtended by the image of the disk in the observer's sky and $D$ is the distance of the observer from the source. $I_{\rm o} = g^3 I_{\rm e}$ follows from Liouville's theorem, where $g = \nu_{\rm o}/\nu_{\rm e}$ is the redshift factor, $\nu_{\rm o}$ is the photon frequency as measured by the distant observer, and $\nu_{\rm e}$ is the photon frequency in the rest frame of the emitter. $r_{\rm e}$ is the emission radius in the disk and $\vartheta_{\rm e}$ is the emission angle (which can be different from the viewing angle of the observer $i$ because of the effect of light bending).

Introducing the transfer function $f$, the observed flux can be rewritten as
\be\label{eq-Fobs}
F_{\rm o} (\nu_{\rm o}) 
= \frac{1}{D^2} \int_{r_{\rm in}}^{r_{\rm out}} \int_0^1
\pi r_{\rm e} \frac{ g^2}{\sqrt{g^* (1 - g^*)}} f(g^*,r_{\rm e},i)
I_{\rm e}(\nu_{\rm e},r_{\rm e},\vartheta_{\rm e}) \, dg^* \, dr_{\rm e} \, ,
\ee
where $r_{\rm in}$ and $r_{\rm out}$ are, respectively, the inner and the outer edge of the accretion disk. In the Novikov-Thorne model~\citep{nt1,nt2}, the inner edge of the disk is assumed to be located at the innermost stable circular orbit (ISCO). The outer edge can be set at some large radius where the emission becomes negligible; in our calculation it will be located at $\sim 1000$. The expression of the transfer function $f$ is~\citep{cun75} 
\be\label{eq-trf}
f(g^*,r_{\rm e},i) = \frac{1}{\pi r_{\rm e}} g 
\sqrt{g^* (1 - g^*)} \left| \frac{\partial \left(X,Y\right)}{\partial \left(g^*,r_{\rm e}\right)} \right| \, ,
\ee
where the relative redshift factor $g^*$ is defined as
\be
g^* = \frac{g - g_{\rm min}}{g_{\rm max} - g_{\rm min}} \, ,
\ee
which ranges from 0 to 1. Here $g_{\rm max}=g_{\rm max}(r_{\rm e},i)$ and $g_{\rm min}=g_{\rm min}(r_{\rm e},i)$ are, respectively, the maximum and the minimum values of the redshift factor $g$ for the photons emitted from the radial coordinate $r_{\rm e}$ and detected by a distant observer with polar coordinate $i$. $\left| \partial \left(X,Y\right)/\partial \left(g^*,r_{\rm e}\right) \right|$ is the Jacobian. The transfer function thus acts as an integration kernel to calculate the spectrum detected by the distant observer starting from the local spectrum at any point of the disk. Let us note that in the specific intensity $I_{\rm e}$, $\nu_{\rm e}$ and $\vartheta_{\rm e}$ must be written in terms of $g^*$ and $r_{\rm e}$. In our model, only the primary image of the accretion disk is taken into account; that is, we neglect secondary and higher order images generated by photons crossing the equatorial plane and then landing on the disk.

\begin{figure}[t]
\begin{center}
\includegraphics[width=10cm]{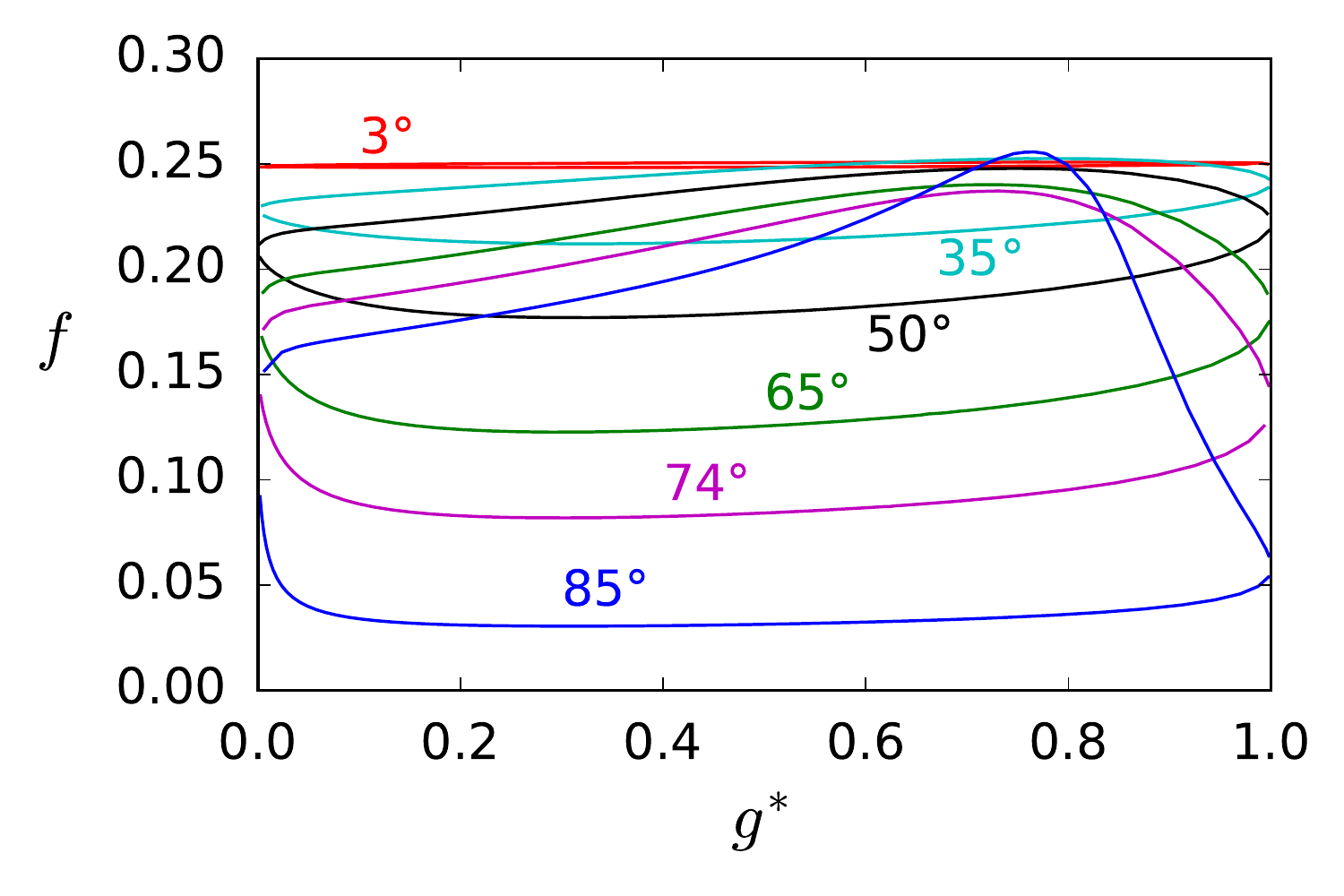}
\end{center}
\vspace{-0.7cm}
\caption{Impact of the viewing angle $i$ on the transfer function $f$. Here the spacetime is described by the Kerr metric with the spin parameter $a_* = 0.998$ and the emission radius is $r_{\rm e} = 4$. The values of the viewing angle are indicated.}
\label{f-trf1}
\vspace{1.0cm}
\begin{center}
\includegraphics[width=10cm]{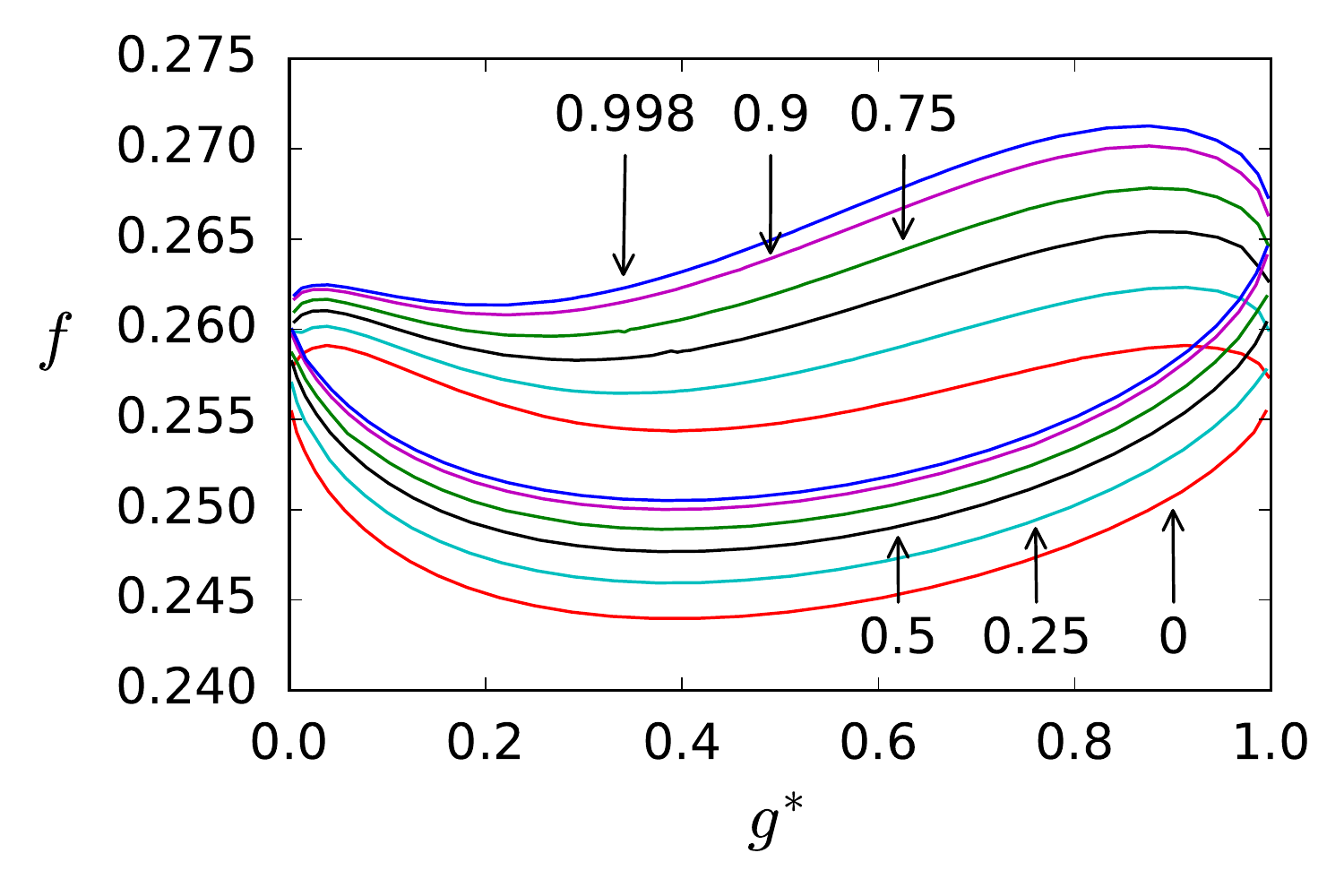}
\end{center}
\vspace{-0.7cm}
\caption{Impact of the dimensionless spin parameter $a_*$ on the transfer function $f$. Here the spacetime is described by the Kerr metric, the emission radius is $r_{\rm e} = 7$, and the viewing angle is $i = 30^\circ$. The values of the spin parameter are indicated.}
\label{f-trf2}
\vspace{0.5cm}
\end{figure}

The transfer function $f (g^*, r_{\rm e}, i)$ only depends on the metric of the spacetime and the position of the distant observer. It takes into account all the relativistic effects (gravitational redshift, Doppler boosting, light bending). For a fixed emission radius $r_{\rm e}$ and viewing angle $i$, the transfer function is a closed curve parameterized by $g^*$, see Fig.~\ref{f-trf1} and Fig.~\ref{f-trf2}. This is true except in the special cases $i = 0$ and $\pi/2$. There is only one point in the disk for which $g^* = 1$ and only one point for which $g^* = 0$. These points are connected by two curves, so we have two branches of the transfer function, say $f^{(1)} (g^*,r_{\rm e},i)$ and $f^{(2)}(g^*,r_{\rm e},i)$. In the case of isotropic emission ($I_{\rm e}$ independent of $\vartheta_{\rm e}$ and of the emission azimuthal angle) in an axisymmetric system (e.g. no orbiting spots), Eq.~(\ref{eq-Fobs}) can be written as
\be
F_{\rm o} (\nu_{\rm o}) 
&=& \frac{1}{D^2} \int_{r_{\rm in}}^{r_{\rm out}} \int_0^1
\frac{\pi r_{\rm e} \, g^2}{\sqrt{g^* (1 - g^*)}} 
\left[ f^{(1)} (g^*,r_{\rm e},i) + f^{(2)} (g^*,r_{\rm e},i) \right] 
I_{\rm e}(\nu_{\rm e},r_{\rm e}) \, dg^* \, dr_{\rm e} \, . \nonumber\\
\ee
If $I_{\rm e}$ does depend on $\vartheta_{\rm e}$, it is necessary to perform the integral twice, one for the upper branch, one for the lower one, so Eq.~(\ref{eq-Fobs}) becomes
\be\label{eq-trf-1-2-b}
\hspace{-1.2cm}
F_{\rm o} (\nu_{\rm o}) 
&=& \frac{1}{D^2} \int_{r_{\rm in}}^{r_{\rm out}} \int_0^1
\frac{\pi r_{\rm e} \, g^2}{\sqrt{g^* (1 - g^*)}} 
f^{(1)} (g^*,r_{\rm e},i)
I_{\rm e}(\nu_{\rm e},r_{\rm e},\vartheta_{\rm e}^{(1)}) \, dg^* \, dr_{\rm e} \nonumber\\
&& + \frac{1}{D^2} \int_{r_{\rm in}}^{r_{\rm out}} \int_0^1
\frac{\pi r_{\rm e} \, g^2}{\sqrt{g^* (1 - g^*)}} 
f^{(2)} (g^*,r_{\rm e},i)
I_{\rm e}(\nu_{\rm e},r_{\rm e},\vartheta_{\rm e}^{(2)}) \, dg^* \, dr_{\rm e} \, ,
\ee
where $\vartheta_{\rm e}^{(1)}$ and $\vartheta_{\rm e}^{(2)}$ indicate the emission angles with relative redshift factor $g^*$, respectively in the branches 1 and 2.

%%%%%%%%%%%%%%%%%%%%%%%%%%%%%%%

\section{Non-Kerr model}\label{s-6}

Model-independent tests of the Kerr metric can be performed by adopting a background more general than the Kerr solution and that includes the Kerr solution as a special case. In addition to the mass $M$ and the spin angular momentum $J$, the metric has a number of deformation parameters used to quantify possible deviations from the Kerr spacetime. The values of these deformation parameters can be constrained by observations. If astrophysical black holes are Kerr black holes, observations should require vanishing deformation parameters. If observations require that at least one of the deformation parameters is non-vanishing, this may be interpreted as an indication of the presence of new physics.

Our current code adopts the Johannsen metric with four deformation parameters~\citep{j-m}. In Boyer-Lindquist coordinates, the line element reads
\be
ds^2 &=& - \frac{\tilde{\Sigma} \left(\Delta - a^2 A_2^2 \sin^2\theta\right)}{B^2} \, dt^2
- \frac{2 a \left[ \left(r^2 + a^2\right) A_1 A_2 - \Delta\right] 
\tilde{\Sigma} \sin^2\theta}{B^2} \, dt \, d\phi \nonumber\\ 
&& + \frac{\tilde{\Sigma}}{\Delta A_5} \, dr^2
+ \tilde{\Sigma} \, d\theta^2
+ \frac{\left[ \left(r^2 + a^2\right)^2 A_1^2 - a^2 \Delta \sin^2\theta\right] 
\tilde{\Sigma} \sin^2\theta}{B^2} \, d\phi^2
\ee
where $a = J/M$,
\be
&&B = \left(r^2 + a^2\right) A_1 - a^2 A_2 \sin^2\theta \, , \quad
\tilde{\Sigma} = \Sigma + f \, , \nonumber\\
&&\Sigma = r^2 + a^2 \cos^2\theta \, , \quad
\Delta = r^2 - 2 M r + a^2 \, ,
\ee
and
\be\label{eq-def}
&& f = \epsilon_3 \frac{M^3}{r} \, , \qquad
A_1 = 1 + \alpha_{13} \left(\frac{M}{r}\right)^3 \, , \nonumber\\
&& A_2 = 1 + \alpha_{22} \left(\frac{M}{r}\right)^2 \, , \qquad
A_5 = 1 + \alpha_{52} \left(\frac{M}{r}\right)^2 \, .
\ee
The deformation parameters are $\epsilon_3$, $\alpha_{13}$, $\alpha_{22}$, and $\alpha_{52}$ and are dimensionless. Such a metric has the correct Newtonian limit and is consistent with the current PPN constraints~\citep{j-m}. It exactly reduces to the Kerr metric for $\epsilon_3 = \alpha_{13} = \alpha_{22} = \alpha_{52} = 0$.

The Johannsen metric has also a Carter-like constant. The normal of the disk at the point of emission becomes (see Appendix~\ref{s-3} for more details)
\be
n^\mu = \left[0, 0, \left(r_{\rm e}^2 + \epsilon_3 \frac{M^3}{r_{\rm e}}\right)^{-1/2}, 0\right] \, .
\ee 
The emission angle $\vartheta_{\rm e}$ can now be written as
\be
\cos \vartheta_{\rm e} = q g \left(r_{\rm e}^2 + \epsilon_3 \frac{M^3}{r_{\rm e}}\right)^{-1/2} \, ,
\ee
where $q^2 = \mathcal{Q}/E^2$, $\mathcal{Q}$ is the Carter-like constant of the photon, and $E$ is the photon energy. In the Johannsen metric, the Carter-like constant has the same form as the Carter constant in the Kerr metric even for non-vanishing deformation parameters~\citep{j-carter}, and $k_\theta = q k_t$ when the photon hits the disk in the equatorial plane. $q$ can be inferred from the photon initial conditions (as in the Kerr metric)
\be
X_0 = \frac{\lambda}{\sin i} \, , \qquad
Y_0 = \sqrt{q^2 + a^2 \cos^2i - \lambda^2\cot^2 i} \, ,
\ee
where $\lambda = L_z/E$.

In the Kerr metric, we have an exterior regular spacetime for $|a| \le M$, which is the condition for the existence of an event horizon. For $|a| > M$, the spacetime has a naked singularity. In the Johannsen metric, if we require a regular exterior region (no singularities or closed time-like curves) we have the following conditions on the deformation parameters~\citep{j-m}
\be\label{eq-regularity}
&&\alpha_{13} \, , \,\, \epsilon_3 \ge - \left( \frac{M + \sqrt{M^2 - a^2}}{M} \right)^3 \, , \nonumber\\
&&\alpha_{22} \, , \,\, \alpha_{25} \ge - \left( \frac{M + \sqrt{M^2 - a^2}}{M} \right)^2 \, .
\ee
We impose these conditions on the deformation parameters in order to avoid spacetimes with pathological properties.

%%%%%%%%%%%%%%%%%%%%%%%%%%%%%%%

\section{Numerical method}\label{s-4}

\begin{figure}[t]
\begin{center}
\includegraphics[width=10cm]{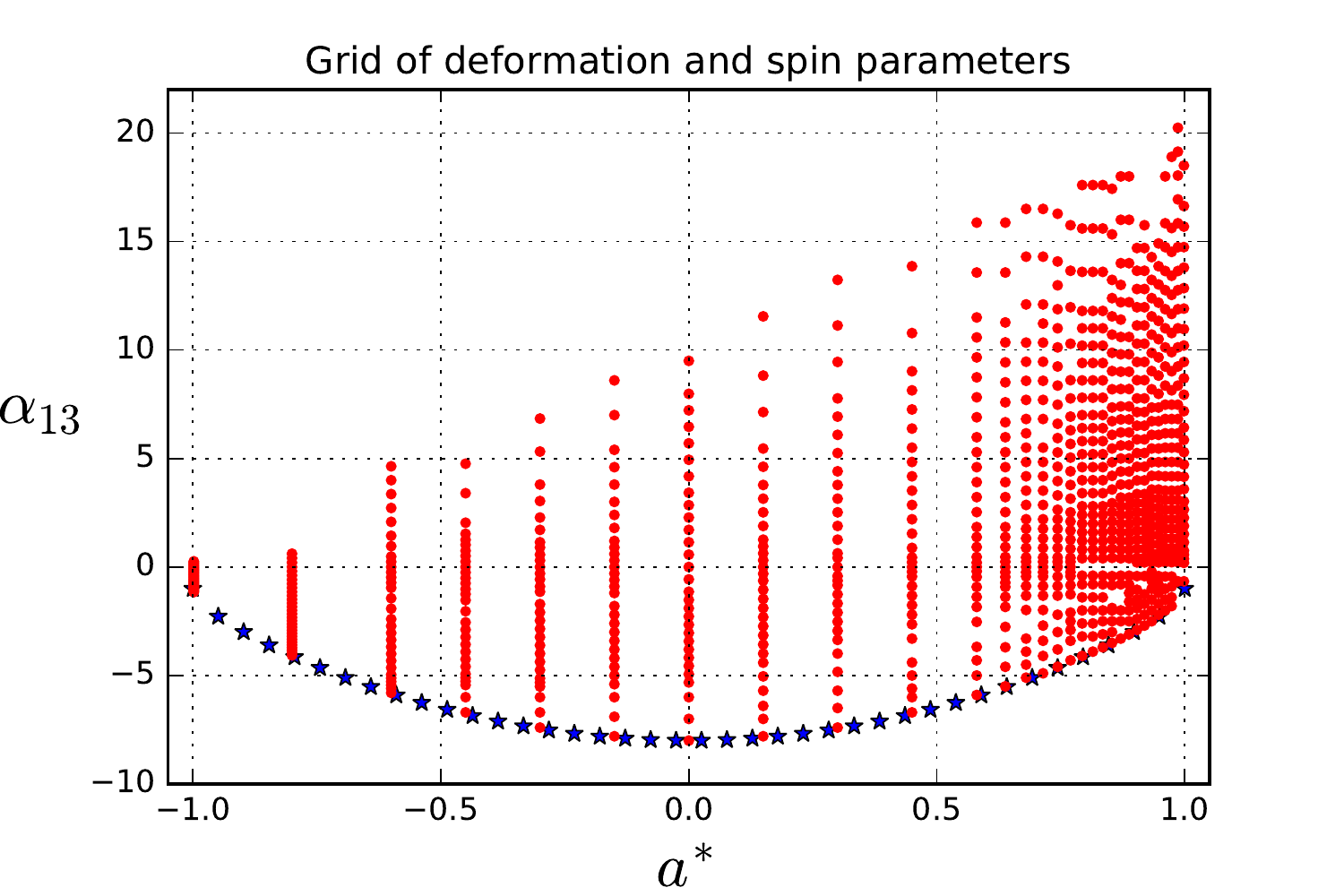}
\end{center}
\caption{Points of the grid of the Master Table for the spin parameter $a_*$ and the deformation parameter $\alpha_{13}$.}
\label{f-grid}
\vspace{0.5cm}
\end{figure}

In this section we describe our algorithm for calculating the transfer function and creating the Master Table that is used to construct the model for {\sc relconv}. The Master Table has data in three dimensions: spin, deformation parameter, and inclination angle. The grid size along each dimension is 30, 30 and 22, respectively. The points along the spin and inclination angle grids are non-uniform and independent of each other. The points along the deformation parameter grid depend on the spin: the points are chosen such that the ISCO radii at each spin, for the range of deformation parameters at that spin, span the range from the minimum Kerr ISCO radius to the maximum Kerr ISCO radius. Fig.~\ref{f-grid} shows these grid points for the deformation parameter $\alpha_{13}$.

At each configuration (namely a grid point with a specific spin, deformation parameter, and inclination angle), we discretize the accretion disk with a grid of 100~emission radii $r_{\rm e}$ and at each emission radius we tabulate the transfer function at 20~equally spaced values of $g^*$ on each branch\footnote{$g^* = 0$ and $g^* = 1$ are replaced, respectively, by $g^* = 0.002$ and $g^* = 0.998$ for numerical reasons, because the Jacobian diverges at $g^* = 0$ and 1.}. The scheme for choosing emission radii and $g^*$ and the values of spin and inclination angles along the grid, are the same as used in the standard {\sc relconv} model. The resulting Master Table has sufficient resolution such that the transfer functions at arbitrary configurations can be interpolated accurately.

Let us notice that the choice of the range of the deformation parameters is somewhat arbitrary. If we adopt the point of view that deviations from the Kerr metric must be small and that the deformation parameters in Eq.~(\ref{eq-def}) are the leading order terms in an expansion, these deformation parameters must also be small and we may restrict the attention to the ranges $(-1,1)$. But here we adopt the same spirit as \citet{j-carter}; we do not impose that the deformation parameters must be small quantities. Since the impact of each deformation parameter on the reflection spectrum is different (see Section~\ref{s-5bis}), it is convenient to adopt different ranges for different deformation parameters.

Before describing each step in detail, we draw an outline of the code. The first step to calculate the transfer function table for each configuration, i.e., for specific values of ($a_*$, $\epsilon_3$, $\alpha_{13}$, $\alpha_{22}$, $\alpha_{52}$, $i$), is to determine the radius of the ISCO. After that, we determine 100 values of the emission radius $r_{\rm e}$ at which we will evaluate the transfer function. We then consider an observer at the distance $D = 10^8$~$M$, so that the spacetime near the observer can be assumed to be flat. We create a grid in the observer plane and, from each point of this grid, we fire photons and calculate their trajectories backward in time from the point of detection in the image plane to the point of emission on the disk. We adjust the position of the photons in the grid adaptively such that they arrive precisely at the radius $r_{\rm e}$ of interest in the accretion disk. We denote by {\it central photon} each photon that hits the accretion disk at one of the target emission radii. We then evaluate the redshift factor $g$ and the emission angle $\vartheta_{\rm e}$ for each central photon. At this point, we fire four photons in a small grid around the central photon and evaluate the Jacobian at each central photon. The size of this small grid is chosen to ensure that the value of the Jacobian has converged and does not change for any smaller grid size. We also use an adaptive method to determine the minimum and maximum values of the redshift factor for each radius $r_{\rm e}$. After finding enough central photons to produce a transfer function curve, we split the central photons in two branches according to their position on the grid relative to the photons associated with the minimum and the maximum redshift factors. Subsequently, we calculate $g^*$ and the transfer function, as defined in Eq.~(\ref{eq-trf}) at each central photon. Since the transfer function at each branch is evaluated at 20 equally spaced values of $g^*$, we interpolate our transfer functions as a function of $g^*$ and obtain their values at the requisite $g^*$ values. This is performed for each of the 100 emission radii of interest. We repeat this process for any configuration ($a_*$, $\epsilon_3$, $\alpha_{13}$, $\alpha_{22}$, $\alpha_{52}$, $i$).

The grid of the plane of the distant observer is adaptive, based on a standard elliptical grid. The points in the grid are defined as
\be\label{defobsgrid}
X_0 (r, \phi) &=& r \cos{\phi}, \nonumber\\
Y_0 (r, \phi) &=& r \sin{\phi} \cos i \, , 
\ee
where $i$ is again the inclination angle of the disk with respect to the line of sight of the distant observer and
\be
\phi &=& \frac{2\pi}{N} j 
\quad j = \{1, 2, ..., N\} \, , \label{eq:defphigrid}\\
r &=& (r_{\rm e})_k 
\quad k = \{1, 2, ..., 100\} \, . \label{eq:defrgrid}
\ee
$N$ is chosen to be 61. The photon trajectories are calculated from the image plane of the distant observer to the emission point in the disk by solving the geodesic equations with the ray-tracing code of~\citet{code-cfm}, which employs an adaptive step-size fourth-order Runge-Kutta-Nystr\"om algorithm~\citep{lund09}. The Christoffel symbols appearing in the geodesic equations are evaluated from their analytical formulas, which have been implemented in the code. Due to gravitational bending, the initial grid of photons does not always hit the accretion disk at the requisite radii. Therefore, the code adjusts $r$ until the photon hits the accretion disk at the emission radius of interest with a precision of $10^{-6}$.

Around each central photon, we choose four photons, whose location on the observer grid relative to the central photon is $(X_0 \pm \Delta X, Y_0 \pm \Delta Y)$, where
\be\label{defobsgrid}
\Delta X = 10^{-4} + 10^{-4} \, X_0 \, , \quad
\Delta Y = 10^{-4} + 10^{-4} \, Y_0 \, . 
\ee
The Jacobian in the expression of the transfer function is calculated at each central photon from 
\begin{align}\label{eq:jacob}
\left| \frac{\partial \left(X,Y\right)}{\partial \left(g^*,r_{\rm e}\right)} \right|
= \left| 
\frac{\partial r_{\rm e}}{\partial X}\frac{\partial g^*}{\partial Y}
- \frac{\partial r_{\rm e}}{\partial Y}\frac{\partial g^*}{\partial X} \right| ^{-1} \, .
\end{align}

Using the preliminary $\phi$ grid defined in Eq.~(\ref{eq:defphigrid}), we find the minimum and maximum redshift factor, $g_{\rm min}$ and $g_{\rm max}$ respectively, for any specific emission radius $r_{\rm e}$. While evaluating the central photons along the $\phi$ grid, we record that central photon as our preliminary $g_{\rm min}$ ($g_{\rm max}$) which has the smallest (largest) redshift factor among the central photons on the grid. We then evaluate the redshift factor on either side of these preliminary extrema with adaptive step-size to move towards the actual $g_{\rm min}$ ($g_{\rm max}$). For each of the extrema, when the change in redshift factor between two consecutive steps is below $10^{-6}$, we assign that central photon as describing those extrema. Using $g_{\rm min}$ and $g_{\rm max}$, we then calculate $g^*$ at every other central photon. Let us denote the $\phi$ values corresponding to $g_{\rm min}$ and $g_{\rm max}$ as $\phi_{\rm min}$ and $\phi_{\rm max}$, respectively. $\phi_{\rm min}$ and $\phi_{\rm max}$ then divide the whole range of $\phi$ (from 0 to $2\pi$) in two branches:
\be\label{eq:branchone}
\phi_{\rm min} < \phi < \phi_{\rm max}
\ee
and
\be\label{eq:branchtwo}
\phi_{\rm min} > \phi > \phi_{\rm max} \, .
\ee

Due to strong gravitational bending, especially at emission radii near the ISCO, it may happen that $\phi_{\rm min}$ and $\phi_{\rm max}$ are close to each other. In this case, the preliminary $\phi$ grid is unable to provide enough central photons on both branches. Consequently, an interpolation of the transfer function can perform poorly. To avoid this, we compare  $g^*$ for each consecutive pair of central photons on the initial $\phi$ grid. If the difference between consecutive $g^*$'s is larger than 0.05, we find additional central photons between the two, such that there are enough $g^*$ to obtain a good interpolation.

The above procedure is repeated for each configuration ($a_*$, $\epsilon_3$, $\alpha_{13}$, $\alpha_{22}$, $\alpha_{52}$, $i$). The data obtained are then fed into a Python routine. For each configuration and each emission radius $r_{\rm e}$, the Python routine splits the data in two branches according to Eqs.~(\ref{eq:branchone}) and~(\ref{eq:branchtwo}), performs a linear interpolation, and generates a pair of transfer functions at constantly spaced $g^*$. Additionally, the emission angles at central photons are also interpolated in the same way to obtain their values at the requisite $g^*$. For each configuration, the data, which comprise of the values of $r_{\rm e}$, $g_{\rm min}$, $g_{\rm max}$, transfer functions, and emission angles $\vartheta_{\rm e}$, are stored in a list and a FITS file (Master Table) is generated with all the configurations.

%%%%%%%%%%%%%%%%%%%%%%%%%%%%%%%

\begin{table}
 \centering
 \caption{}
\begin{tabular}{|c|cccccc|}
\hline
&&&& $g^*$ &&\\
$r_{\rm e} = 1.2468$ && 2 & 7 & 11 & 15 & 19 \\
\hline
This work & $g$ & 0.06573 & 0.19584 & 0.29993 & 0.40401 & 0.50810 \\
& $f^{(1)}$ & 0.10682 & 0.12469 & 0.12479 & 0.11934 & 0.10365 \\ 
& $f^{(2)}$ & 0.02651 & 0.02935 & 0.03863 & 0.05128 & 0.07216 \\
\hline
{\sc relline} & $g$ & 0.06571 & 0.19568 & 0.29965 & 0.40363 & 0.50761 \\
& $f^{(1)}$ & 0.10729 & 0.12462 & 0.12473 & 0.11897 & 0.10385 \\
& $f^{(2)}$ & 0.02639 & 0.02929 & 0.03862 & 0.05122 & 0.07188 \\
\hline
\hline
&&&& $g^*$ &&\\
$r_{\rm e} = 4.7197$ && 2 & 7 & 11 & 15 & 19 \\
\hline
This work & $g$ & 0.48405 & 0.71230 & 0.89491 & 1.07752 & 1.26013 \\
& $f^{(1)}$ & 0.18295 & 0.19639 & 0.21602 & 0.23010 & 0.18632 \\
& $f^{(2)}$ & 0.11339 & 0.09843 & 0.10058 & 0.10676 & 0.12315 \\
\hline
{\sc relline} & $g$ & 0.48406 & 0.71229 & 0.89486 & 1.07745 & 1.26003 \\
& $f^{(1)}$ & 0.18286 & 0.19636 & 0.21598 & 0.23007 & 0.18653 \\ 
& $f^{(2)}$ & 0.11314 & 0.09837 & 0.10054 & 0.10672 & 0.12305 \\
\hline
\hline
&&&& $g^*$ &&\\
$r_{\rm e} = 41.309$ && 2 & 7 & 11 & 15 & 19 \\
\hline
This work & $g$ & 0.85331 & 0.93102 & 0.99318 & 1.05535 & 1.11751 \\
& $f^{(1)}$ & 0.11711 & 0.12170 & 0.12347 & 0.13112 & 0.11257 \\
& $f^{(2)}$ & 0.10404 & 0.10249 & 0.10245 & 0.10287 & 0.10442 \\
\hline
{\sc relline} & $g$ & 0.85331 & 0.93102 & 0.99318 & 1.05535 & 1.11751 \\
& $f^{(1)}$ & 0.11710 & 0.12169 & 0.12344 & 0.13129 & 0.11258 \\
& $f^{(2)}$ & 0.10403 & 0.10248 & 0.10244 & 0.10287 & 0.10440 \\
\hline
\end{tabular}
\vspace{0.3cm}
\tablenotetext{0}{Comparison between the redshift factors $g$ and the transfer functions calculated by the code described in this Paper and by {\sc relline} at three different values of the emission radius $r_{\rm e}$ and five values of $g^*$ [the Master Table has 20 equally spaced values of $g^*$ with $g^*(1) = 0.002$ and $g^*(20) = 0.998$, see the text for details]. These quantities are calculated for the Kerr metric with the spin parameter $a_* = 0.9982$ and the cosine of the viewing angle $\mu = 0.3221819$ (viewing angle $i \approx 71^\circ$). For every $g^*$, there are two values for the transfer function, corresponding, respectively, to the values of the transfer function in the upper (first line) and lower (second line) branches.} 
\label{tab1}
\end{table}

\begin{table}
 \centering
 \caption{}
\begin{tabular}{|c|cccccc|}
\hline
&&&& $g^*$ &&\\
$r_{\rm e} = 1.2468$ && 2 & 7 & 11 & 15 & 19 \\
\hline
This work & $\cos\vartheta_{\rm e}^{(1)}$ & 0.18493 & 0.38686 & 0.45599 & 0.4796 & 0.44072 \\ 
& $\cos\vartheta_{\rm e}^{(2)}$ & 0.02857 & 0.06524 & 0.11042 & 0.17416 & 0.28502 \\
\hline
{\sc relline} & $\cos\vartheta_{\rm e}^{(1)}$ & 0.18712 & 0.38670 & 0.45563 & 0.47917 & 0.44027 \\
& $\cos\vartheta_{\rm e}^{(2)}$ & 0.02856 & 0.06510 & 0.110244 & 0.17391 & 0.28299 \\
\hline
\hline
&&&& $g^*$ &&\\
$r_{\rm e} = 4.7197$ && 2 & 7 & 11 & 15 & 19 \\
\hline
This work & $\cos\vartheta_{\rm e}^{(1)}$ & 0.50880 & 0.88950 & 0.92303 & 0.81484 & 0.62527 \\
& $\cos\vartheta_{\rm e}^{(2)}$ & 0.17497 & 0.23108 & 0.29274 & 0.36383 & 0.46190 \\
\hline
{\sc relline} & $\cos\vartheta_{\rm e}^{(1)}$ & 0.51160 & 0.88938 & 0.92291 & 0.81471 & 0.62541 \\
& $\cos\vartheta_{\rm e}^{(2)}$ & 0.17485 & 0.23094 & 0.29263 & 0.36369 & 0.46160 \\
\hline
\hline
&&&& $g^*$ &&\\
$r_{\rm e} = 41.309$ && 2 & 7 & 11 & 15 & 19 \\
\hline
This work & $\cos\vartheta_{\rm e}^{(1)}$ & 0.29837 & 0.46283 & 0.48959 & 0.41144 & 0.38035 \\
& $\cos\vartheta_{\rm e}^{(2)}$ & 0.27747 & 0.30030 & 0.32028 & 0.34107 & 0.36408 \\
\hline
{\sc relline} & $\cos\vartheta_{\rm e}^{(1)}$ & 0.29831 & 0.46284 & 0.48977 & 0.41098 & 0.38034 \\
& $\cos\vartheta_{\rm e}^{(2)}$ & 0.27746 & 0.30028 & 0.32026 & 0.34105 & 0.36404 \\
\hline
\end{tabular} 
\vspace{0.3cm}
\tablenotetext{0}{As in Tab.~\ref{tab1} for the cosine of the emission angle. $\vartheta_{\rm e}^{(1)}$ and $\vartheta_{\rm e}^{(2)}$ refer, respectively, to the upper and lower branches.}
\label{tab2}
\end{table}

\begin{table}
 \centering
 \caption{}
\begin{tabular}{|c|cccccc|}
\hline
&&&& $g^*$ &&\\
$r_{\rm e} = 7.5154$ && 2 & 7 & 11 & 15 & 19 \\
\hline
This work & $g$ & 0.64784 & 0.73789 & 0.80994 & 0.88199 & 0.95403 \\
& $f^{(1)}$ & 0.25996 & 0.25413 & 0.25311 & 0.25484 & 0.25644 \\ 
& $f^{(2)}$ & 0.24950 & 0.24350 & 0.24375 & 0.24604 & 0.25107 \\
\hline
{\sc relline} & $g$ & 0.64784 & 0.73789 & 0.80994 & 0.88198 & 0.95402 \\
& $f^{(1)}$ & 0.25997 & 0.25414 & 0.25309 & 0.25482 & 0.25644 \\
& $f^{(2)}$ & 0.24946 & 0.24347 & 0.24372 & 0.24601 & 0.25099 \\
\hline
\hline
&&&& $g^*$ &&\\
$r_{\rm e} = 25.786$ && 2 & 7 & 11 & 15 & 19 \\
\hline
This work & $g$ & 0.86191 & 0.91369 & 0.95511 & 0.99653 & 1.03795 \\
& $f^{(1)}$ & 0.27586 & 0.27202 & 0.27114 & 0.27350 & 0.27544 \\
& $f^{(2)}$ & 0.27185 & 0.26957 & 0.26945 & 0.27018 & 0.27221 \\
\hline
{\sc relline} & $g$ & 0.86191 & 0.91369 & 0.95511 & 0.99653 & 1.03795 \\
& $f^{(1)}$ & 0.27587 & 0.27199 & 0.27113 & 0.27350 & 0.27545 \\
& $f^{(2)}$ & 0.27184 & 0.26956 & 0.26945 & 0.27016 & 0.27226 \\
\hline
\hline
&&&& $g^*$ &&\\
$r_{\rm e} = 158.52$ && 2 & 7 & 11 & 15 & 19 \\
\hline
This work & $g$ & 0.95632 & 0.97738 & 0.99422 & 1.01106 & 1.02791 \\
& $f^{(1)}$ & 0.27496 & 0.27419 & 0.27388 & 0.27449 & 0.27494 \\
& $f^{(2)}$ & 0.27424 & 0.27385 & 0.27381 & 0.27391 & 0.27427 \\
\hline
{\sc relline} & $g$ & 0.95632 & 0.97738 & 0.99422 & 1.01106 & 1.02791 \\
& $f^{(1)}$ & 0.27487 & 0.27419 & 0.27388 & 0.27449 & 0.27496 \\
& $f^{(2)}$ & 0.27425 & 0.27385 & 0.27381 & 0.27391 & 0.27428 \\
\hline
\end{tabular}
\vspace{0.3cm}
\tablenotetext{0}{As in Tab.~\ref{tab1}, but for the Kerr metric with the spin parameter $a_* = -0.45$ and the cosine of the viewing angle $\mu = 0.8622873$ (viewing angle $i \approx 30^\circ$).}
\label{tab3}
\end{table}

\begin{table}
 \centering
 \caption{}
\begin{tabular}{|c|cccccc|}
\hline
&&&& $g^*$ &&\\
$r_{\rm e} = 7.5154$ && 2 & 7 & 11 & 15 & 19 \\
\hline
This work & $\cos\vartheta_{\rm e}^{(1)}$ & 0.68396 & 0.85172 & 0.93467 & 0.98443 & 0.99378 \\ 
& $\cos\vartheta_{\rm e}^{(2)}$ & 0.60012 & 0.66383 & 0.73092 & 0.80741 & 0.90622 \\
\hline
{\sc relline} & $\cos\vartheta_{\rm e}^{(1)}$ & 0.68407 & 0.85165 & 0.93461 & 0.98437 & 0.99393 \\
& $\cos\vartheta_{\rm e}^{(2)}$ & 0.60003 & 0.66374 & 0.73079 & 0.80732 & 0.90588 \\
\hline
\hline
&&&& $g^*$ &&\\
$r_{\rm e} = 25.786$ && 2 & 7 & 11 & 15 & 19 \\
\hline
This work & $\cos\vartheta_{\rm e}^{(1)}$ & 0.79098 & 0.87153 & 0.91439 & 0.94004 & 0.94934 \\
& $\cos\vartheta_{\rm e}^{(2)}$ & 0.76046 & 0.79832 & 0.83424 & 0.87305 & 0.91854 \\
\hline
{\sc relline} & $\cos\vartheta_{\rm e}^{(1)}$ & 0.79099 & 0.87160 & 0.91437 & 0.94005 & 0.94944 \\
& $\cos\vartheta_{\rm e}^{(2)}$ & 0.76040 & 0.79829 & 0.83420 & 0.87300 & 0.91847 \\
\hline
\hline
&&&& $g^*$ &&\\
$r_{\rm e} = 158.52$ && 2 & 7 & 11 & 15 & 19 \\
\hline
This work & $\cos\vartheta_{\rm e}^{(1)}$ & 0.83306 & 0.85766 & 0.87355 & 0.88583 & 0.89515 \\
& $\cos\vartheta_{\rm e}^{(2)}$ & 0.82783 & 0.84466 & 0.85909 & 0.87939 & 0.88995 \\
\hline
{\sc relline} & $\cos\vartheta_{\rm e}^{(1)}$ & 0.83307 & 0.85768 & 0.87354 & 0.88584 & 0.89517 \\
& $\cos\vartheta_{\rm e}^{(2)}$ & 0.82782 & 0.84465 & 0.8591 & 0.87398 & 0.88994 \\
\hline
\end{tabular} 
\vspace{0.3cm}
\tablenotetext{0}{As in Tab.~\ref{tab2}, but for the Kerr metric with the spin parameter $a_* = -0.45$ and the cosine of the viewing angle $\mu = 0.8622873$ (viewing angle $i \approx 30^\circ$).}
\label{tab4}
\end{table}

\section{Comparison with existing codes for the Kerr metric}\label{s-5}

In this section we want to test if the lines shapes and transfer functions produced by the code discussed in the Paper agree with existing simulations for the Kerr case. We will use the {\sc relline} model to incorporate the non-Kerr relativistic smearing of the reflection spectrum and therefore this model is also used as comparison. The very good agreement of the {\sc relline} model with other existing model codes has been shown previously \citep[see][]{relline1}.

The transfer functions and additional information of the ray-tracing simulations are stored in a table, which is in the same format as the table used by the {\sc relline} model. This allows us to use the {\sc relline} model code to predict the line shape for the non-Kerr spacetimes and directly compare the calculated shapes and transfer functions for the Kerr case.

Tab.~\ref{tab1} shows the values of the transfer functions from our code and from {\sc relline} for the Kerr metric with the spin parameter $a_* = 0.9982$ and the cosine of the viewing angle $\mu = 0.3221819$ (viewing angle $i = 71.21^\circ$). For illustration, we report three emission radii, namely $r_{\rm e} = 1.2468$, 4.7197, and 41.309, and five relative redshift factors $g^*$. The actual values of the redshift factor, $g$, as computed by the two codes, are also shown. For every $r_{\rm e}$ and $g^*$ there are two values of the transfer function; the first line refers to the values of the transfer function in the first branch, $f^{(1)}$ in Eq.~(\ref{eq-trf-1-2-b}), the second line to the values of the transfer function in the second branch, $f^{(2)}$. Tab.~\ref{tab2} shows the values of the emission angles, $\vartheta_{\rm e}^{(1)}$ and $\vartheta_{\rm e}^{(2)}$, for the same configuration. Tab.~\ref{tab3} and Tab.~\ref{tab4} report the transfer functions and the cosines of the emission angles for the Kerr metric with $a_* = -0.45$ and $\mu = 0.8622873$ (viewing angle $i = 30.43^\circ$).

\begin{figure}[t]
\begin{center}
\includegraphics[width=8cm]{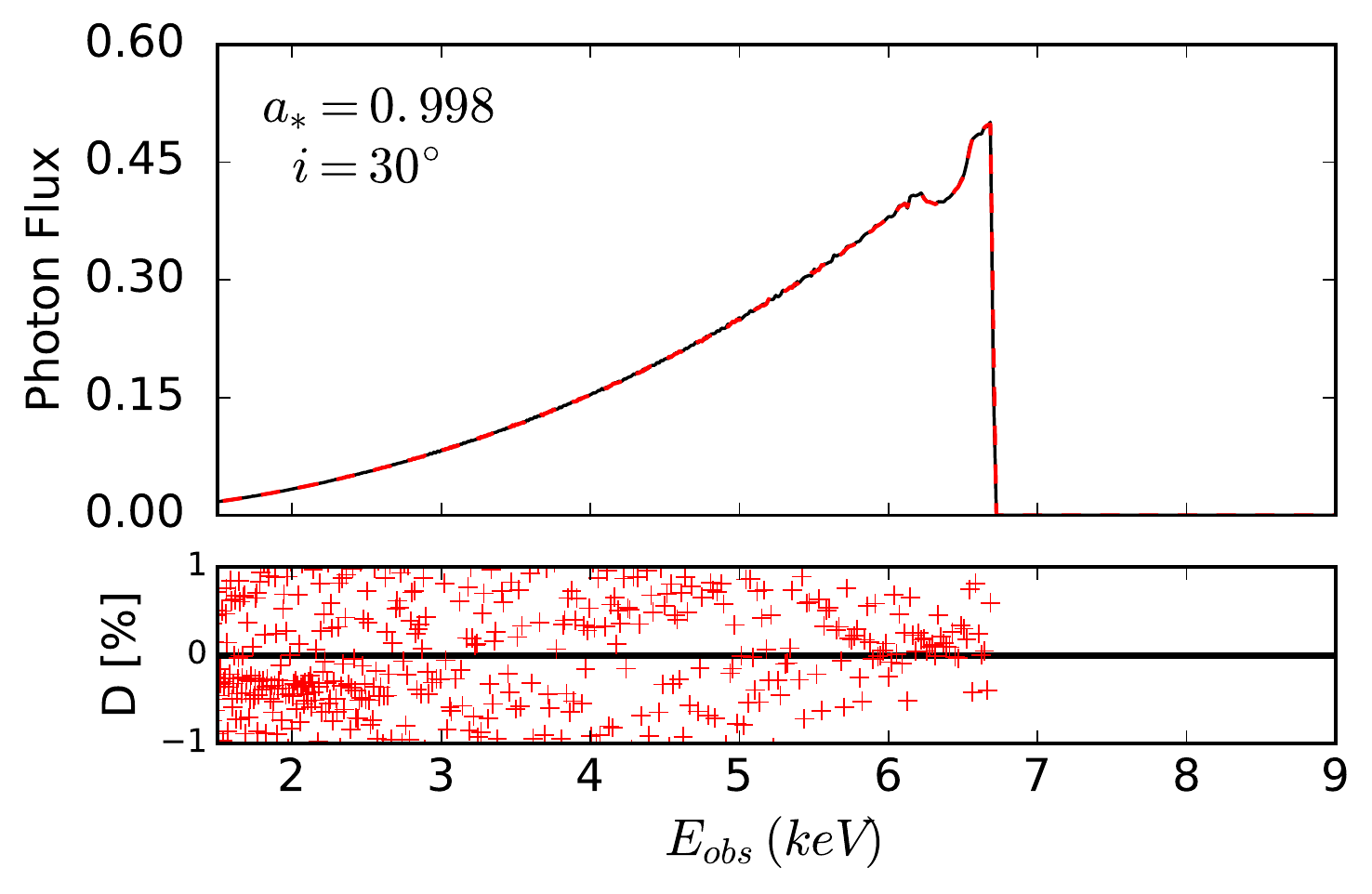}
\includegraphics[width=8cm]{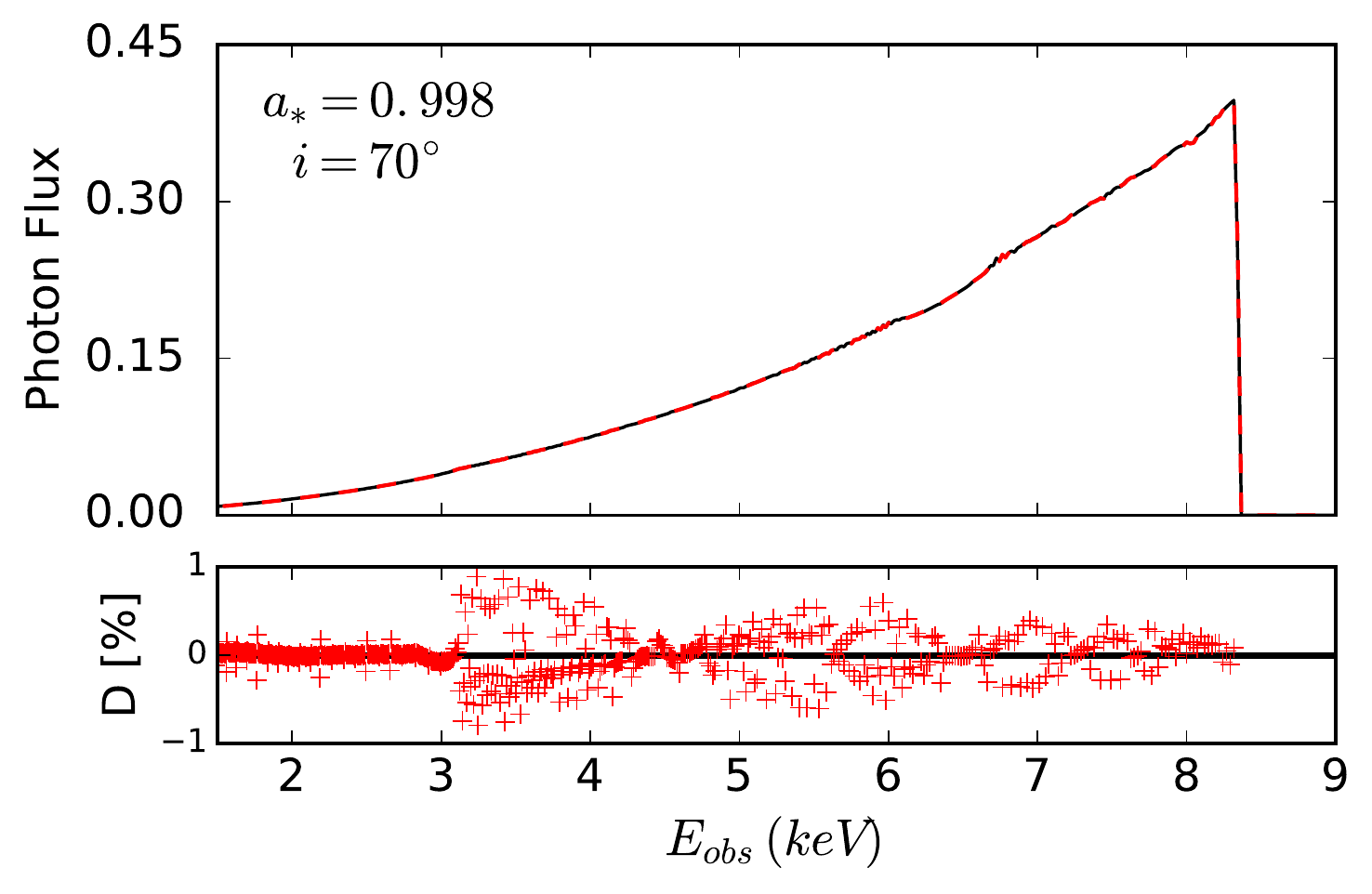} \\ \vspace{0.5cm}
\includegraphics[width=8cm]{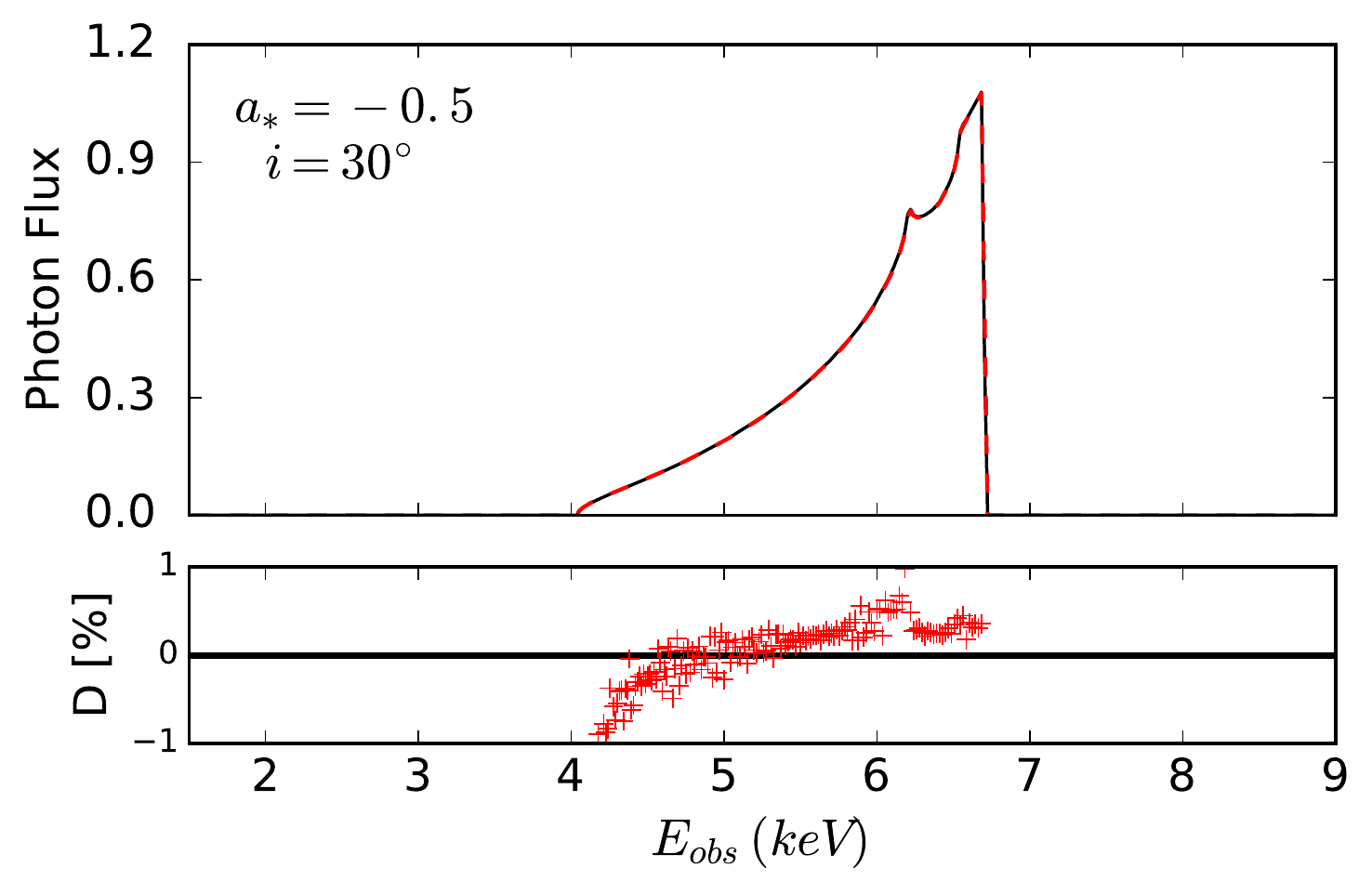}
\includegraphics[width=8cm]{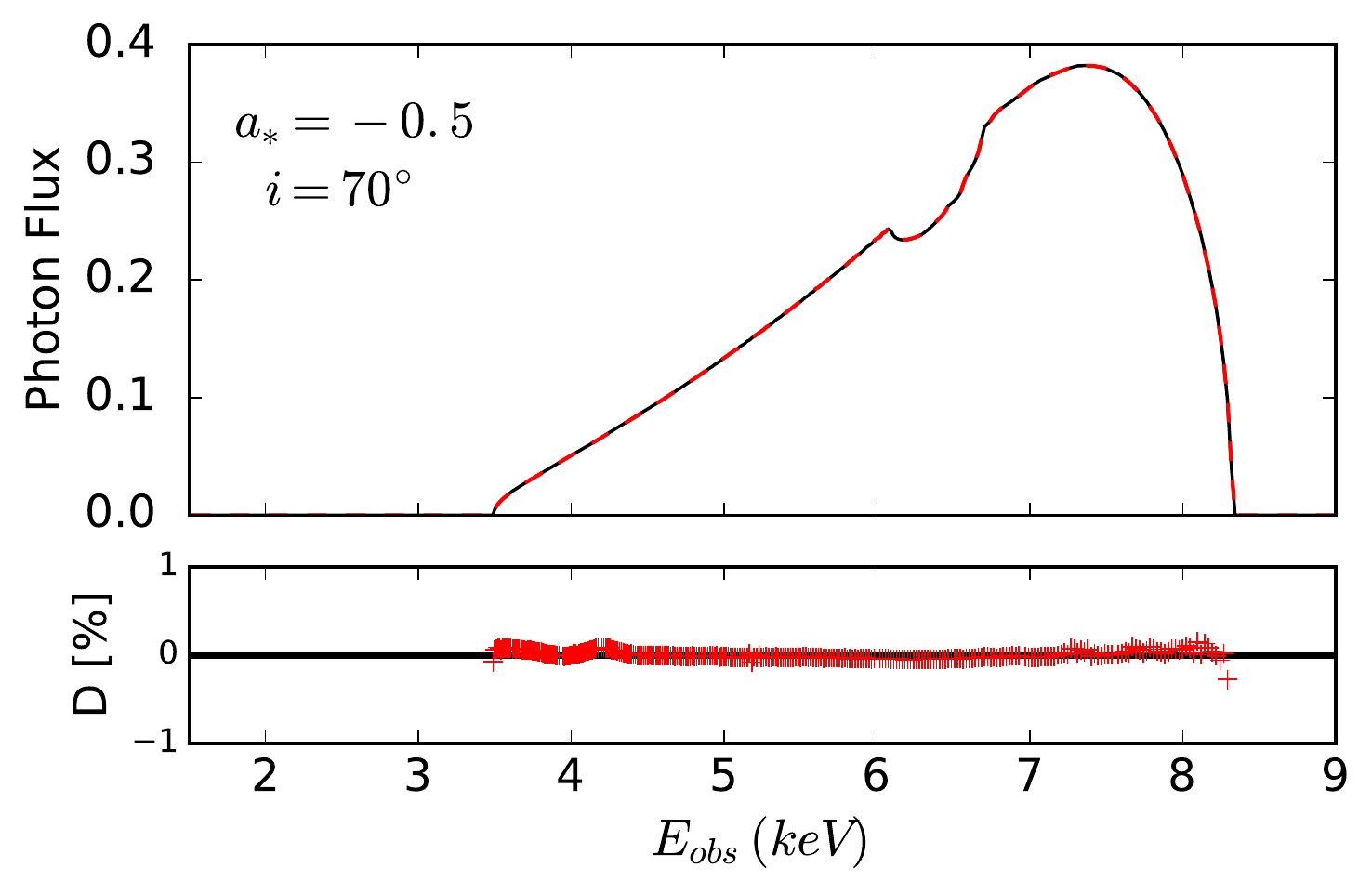}
\end{center}
\caption{Comparison between single iron line shapes in the Kerr metric generated by our new code (red dashed lines) and by {\sc relline} (black lines) for different values of the spin parameter $a_*$ and the viewing angle $i$. The difference in percentage between the two lines at every energy bin is shown in the box below every panel and it is usually within 1\%. Top left panel: $a_* = 0.998$ and $i = 30^\circ$. Top right panel: $a_* = 0.998$ and $i = 70^\circ$. Bottom left panel: $a_* = -0.5$ and $i = 30^\circ$. Bottom right panel: $a_* = -0.5$ and $i = 70^\circ$. }
\label{f-tests}
\vspace{0.5cm}
\end{figure}

In Fig.~\ref{f-tests} we compare the relativistic line between our code and {\sc relline} for a few representative cases. In the top panels, we have a fast-rotating black hole with spin parameter $a_* = 0.998$. In the bottom panels, we have the iron line from a retrograde disk and the black hole spin is $a_* = -0.5$. In the left panels, the viewing angle is $i = 30^\circ$, while it is $i = 70^\circ$ in the right panels. The box below every panel shows the difference in percentage between the two lines. This is usually within 1\%. As can be seen, both line shapes are in very good agreement and therefore we conclude that the presented ray-tracing code agrees with existing model codes for calculating relativistic reflection.

As a last check, we perform a more quantitative analysis to compare the level of accuracy of our transfer functions. We simulate some observations with LAD/eXTP~\citep{extpwp}. eXTP is a future X-ray mission and LAD will have an effective area of 3.4~m$^2$ at 6~keV. The theoretical model is a power law plus an iron line generated by {\sc relline}. We consider the case of a bright binary (flux between 2 and 10~keV at the level of $10^{-9}$~erg/s/cm$^2$) and the equivalent width of the iron line is $\sim 400$~eV. We adopt an exposure time of 1~Ms. All these parameters (brightness of the source, equivalent width of the iron line, exposure time) are quite optimistic, so we can obtain a good measurement. We then fit the simulated data with both our single iron lines for Kerr and with {\sc relline}, and we compare the difference.

\begin{figure}[t]
\begin{center}
\includegraphics[width=10cm]{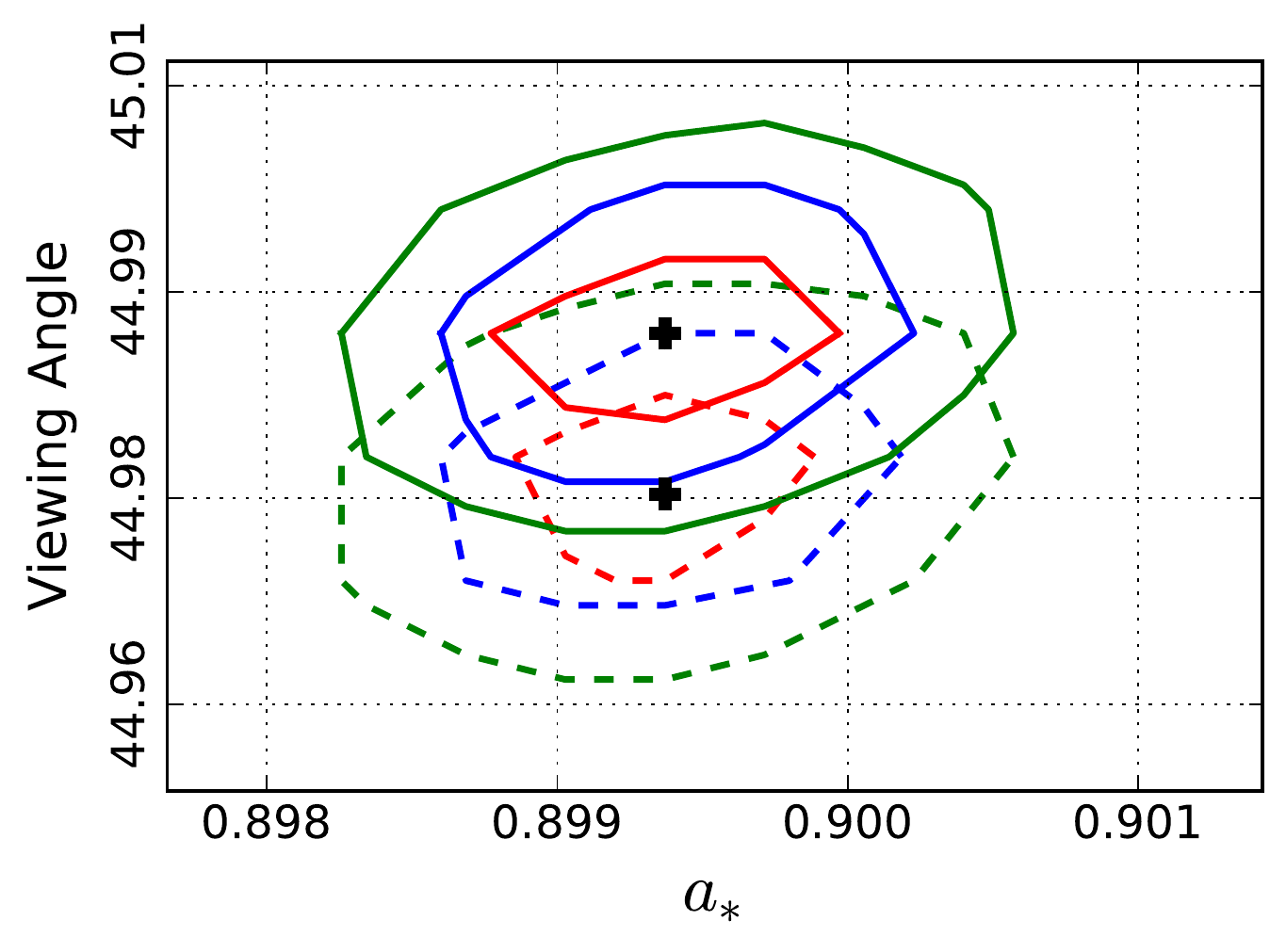}
\end{center}
\caption{$\Delta\chi^2$ contours of the simulations described at the end of Section~\ref{s-5}. The simulated data have been obtained by {\sc relline}, plugging the spin parameter $a_* = 0.9$ and the viewing angle $i = 45^\circ$. The data have been fitted with the iron lines in the Kerr metric generated by the code presented in this work (dashed curves) and with the iron lines generated by {\sc relline} (solid curves). The black cross indicates the position of the minimum of $\chi^2$, while the red, blue, and green curves indicate, respectively, the 1-, 2-, 3-$\sigma$ limits. See the text for more details.}
\label{f-tests2}
\vspace{0.5cm}
\end{figure}

The two models provide measurements in very good agreement, suggesting that our code can compute the transfer function in the Kerr metric with the necessary precision for very accurate measurements. Fig.~\ref{f-tests2} shows the result of one of our simulations. The input spin parameter and the input viewing angle of the simulations are, respectively, $a_* = 0.9$ and $i = 45^\circ$. When we fit the simulated data with the table of transfer functions generated by our new code, we find (here the error is at the 90\% confidence level)
\be
a_* = 0.8996 \pm 0.0008 \, , \quad
i = 44.977^\circ \pm 0.010^\circ \, .
\ee
When we use {\sc relline}, we obtain
\be
a_* = 0.8997 \pm 0.0008 \, , \quad
i = 44.992^\circ \pm 0.010^\circ \, .
\ee
The difference between the two models is much smaller than what one can imagine to measure with the next generation of X-ray satellites (and maybe even with X-ray reflection spectroscopy in general). Fig.~\ref{f-tests2} shows the contour of $\Delta\chi^2$. We have obtained similar results with different input parameters.

%%%%%%%%%%%%%%%%%%%%%%%%%%%%%%%

\section{Single line shapes in Non-Kerr spacetimes}\label{s-5bis}

Unlike existing codes for the Kerr metric, our transfer function code uses formulas valid for any stationary, axisymmetric, and asymptotically flat black hole spacetime. It is sufficient to set the deformation parameters to a non-vanishing value to obtain the corresponding transfer function and single line shape.

\begin{figure}[t]
\begin{center}
\includegraphics[width=14.5cm]{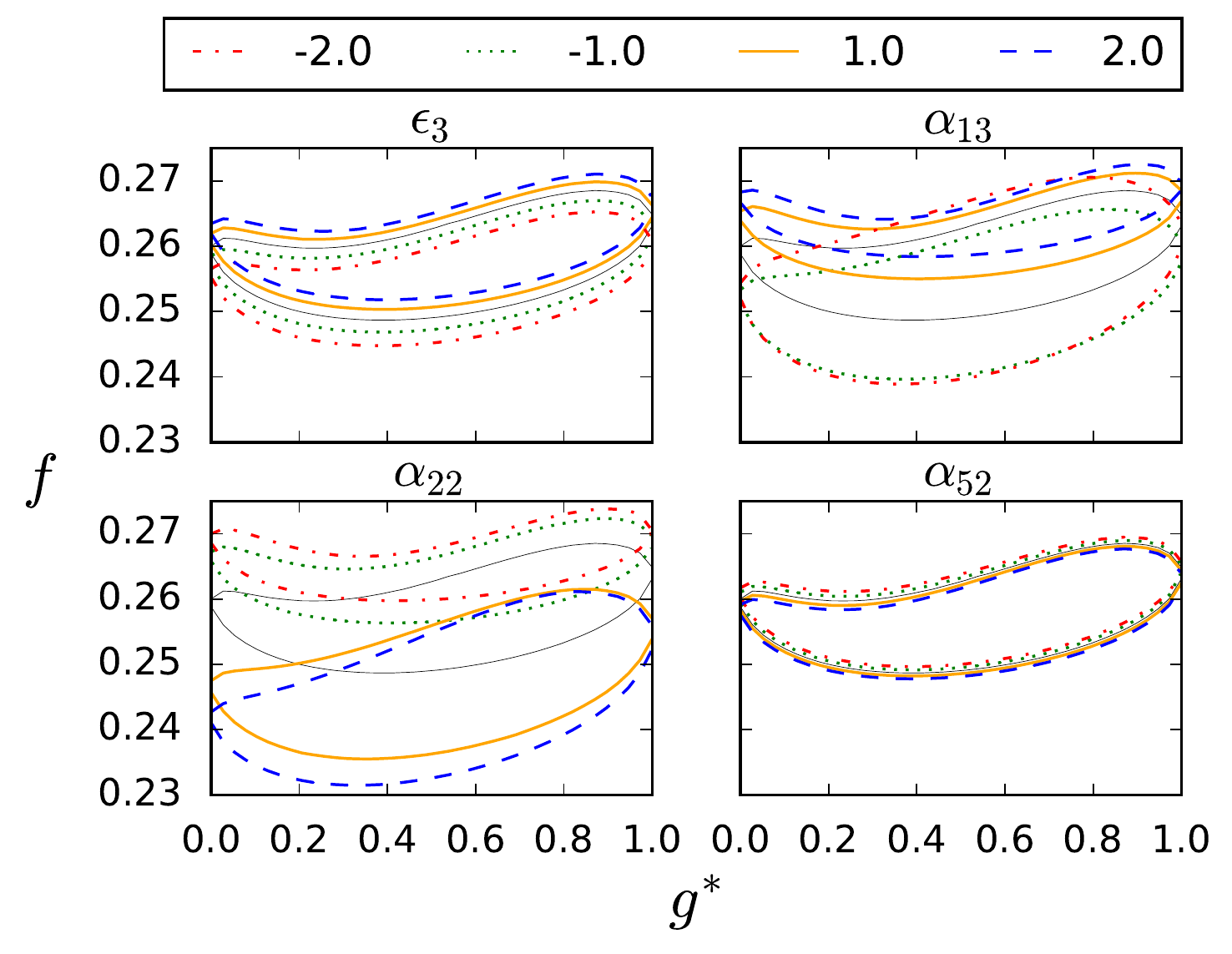}
\end{center}
\caption{Impact of the deformation parameters $\epsilon_3$, $\alpha_{13}$, $\alpha_{22}$, and $\alpha_{52}$ on the transfer function $f$. The spacetime is described by the Johannsen metric with the spin parameter $a_* = 0.8$. The emission radius is $r_{\rm em} = 6.855$ and the viewing angle is $i = 30^\circ$. In every plot, one of the deformation parameters assumes the values 0 (black solid line), $\pm 1$, and $\pm 2$, while the other deformation parameters vanish.}
\label{f-trf-def}
\vspace{0.5cm}
\end{figure}

Examples of transfer functions in Johannsen metric are shown in Fig.~\ref{f-trf-def}. Each panel shows the impact of one of the deformation parameters on the transfer function, assuming that the other three deformation parameters vanish. All the transfer functions have been evaluated at the emission radius $r_{\rm e} = 6.855$, for a viewing angle $i = 30^\circ$, and for a spin parameter $a_*= 0.8$. The transfer function for the Kerr metric with $\epsilon_3 = \alpha_{13} = \alpha_{22} = \alpha_{52} = 0$ is the black solid curve. The other curves correspond to the transfer functions for $\epsilon_3 = \pm 1$ and $\pm 2$ (top left panel), $\alpha_{13} = \pm 1$ and $\pm 2$ (top right panel), $\alpha_{22} = \pm 1$ and $\pm 2$ (bottom left panel), and $\alpha_{52} = \pm 1$ and $\pm 2$ (bottom right panel).

\begin{figure}[t]
\begin{center}
\includegraphics[width=14.5cm]{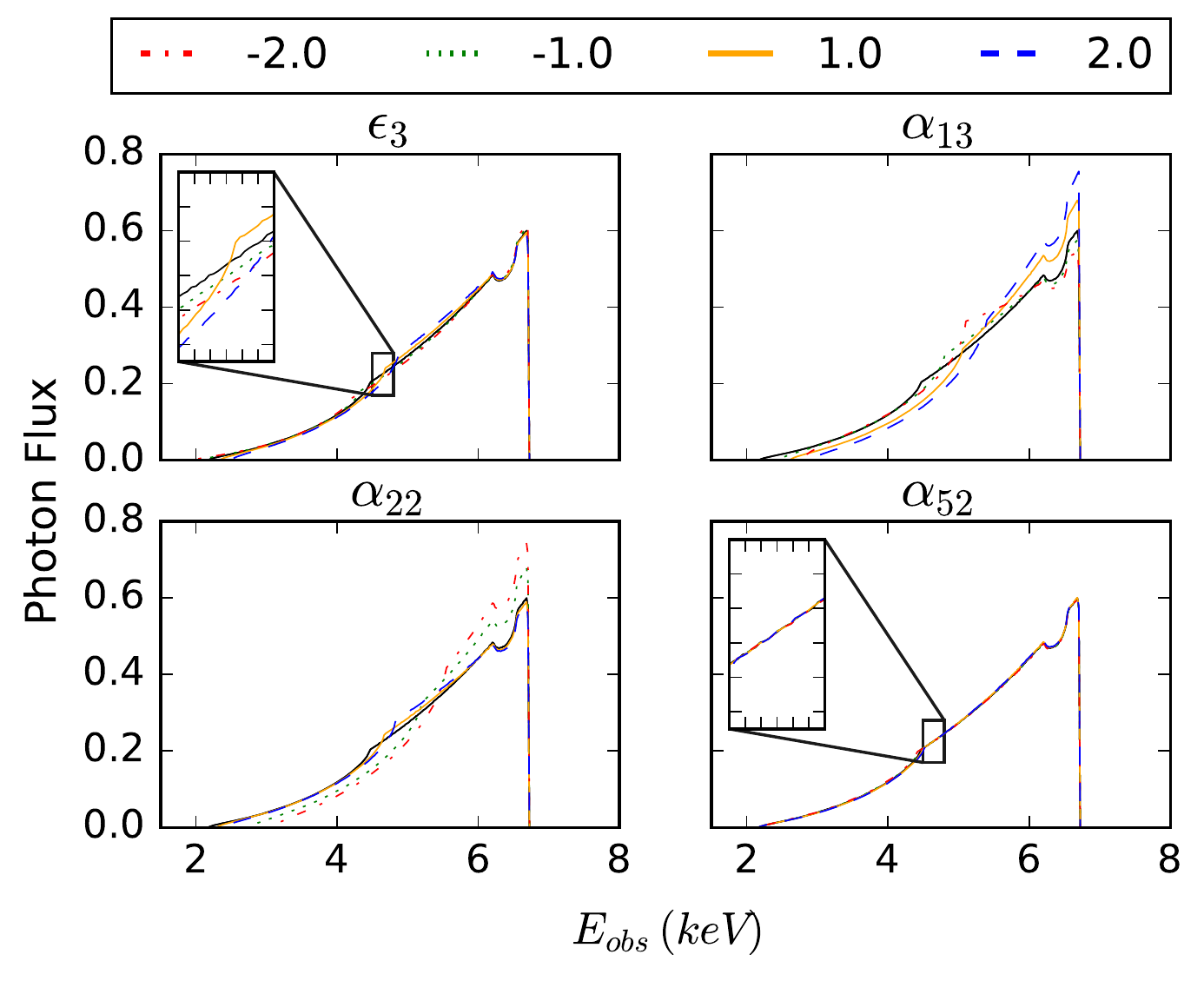}
\end{center}
\caption{Impact of the deformation parameters $\epsilon_3$, $\alpha_{13}$, $\alpha_{22}$, and $\alpha_{52}$ on the iron line shape. The spacetime is described by the Johannsen metric with the spin parameter $a_* = 0.8$. The viewing angle is $i = 30^\circ$. The profile of the emissivity is modeled with a simple power law with emissivity index $q=3$, namely $I_{\rm e} \propto 1/r_{\rm e}^3$. The inner edge of the disk is at the ISCO radius $r_{\rm in} = r_{\rm ISCO}$, and the outer edge is at $r_{\rm out} = 400$.}
\label{f-line-def}
\vspace{0.5cm}
\end{figure}

The single iron line shapes of the spacetimes considered in Fig.~\ref{f-trf-def} are shown in Fig.~\ref{f-line-def}. The emission line is at $E_{\rm e} = 6.4$~keV. The inner edge of the disk is set at the ISCO radius, while the other edge is at $r_{\rm out} = 400$. The local spectrum $I_{\rm e}$ is modeled with a power law with emissivity index equal to 3, namely $I_{\rm e} \propto 1/r^3_{\rm e}$. As already discussed in~\citet{j-m}, $\alpha_{13}$ and $\alpha_{22}$ strongly affect the ISCO radius and the iron line shape, $\epsilon_3$ has a moderate impact on both the ISCO radius and the iron line shape, while $\alpha_{52}$ does not affect the ISCO radius and has an extremely weak impact on the iron line shape.

\begin{figure}[t]
\begin{center}
\includegraphics[width=14.5cm]{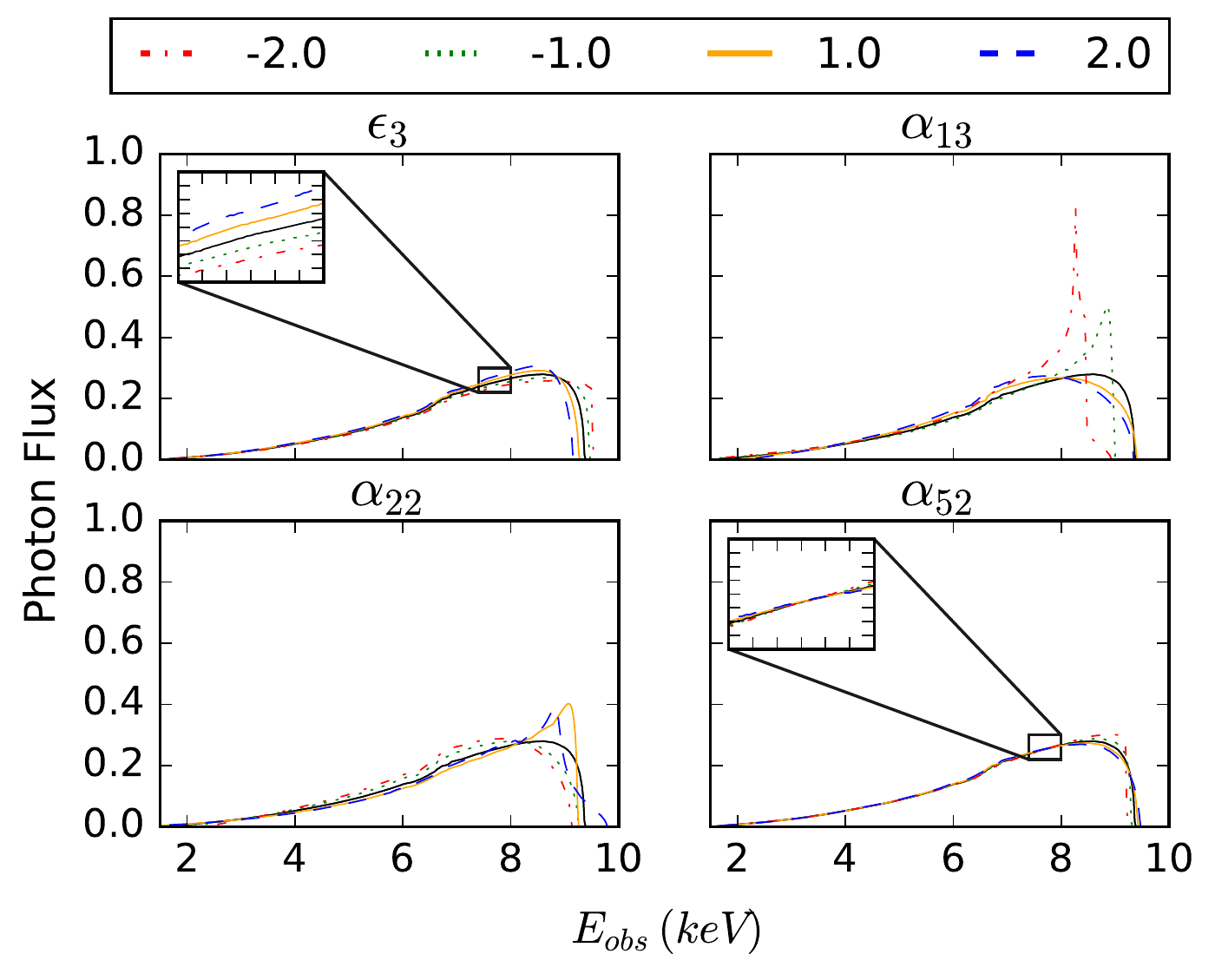}
\end{center}
\caption{As in Fig.~\ref{f-line-def}, but for the viewing angle $i = 80^\circ$.}
\label{f-line-def80}
\vspace{0.5cm}
\end{figure}

In Fig.~\ref{f-line-def}, the maximum energy of the line does not change with the value of the deformation parameter. For $i = 30^\circ$, the Doppler effect is moderate, and the photons with the highest energies come from relatively large radii ($r_{\rm e} \approx 10$-20~$M$). This suggests that the effects of these deformation parameters are localized quite close to the black hole. For larger viewing angles, the Doppler effect is stronger, while the gravitational redshift is the same because it does not depend on $i$. The result is that the photons with the highest energies come from smaller radii. Fig.~\ref{f-line-def80} shows the iron lines in Fig.~\ref{f-line-def} for $i = 80^\circ$. The impact of the deformation parameters is now stronger, and, in particular, the very high energy part of the line does depend on the value of the deformation parameters. Even a non-vanishing $\alpha_{52}$, which had an extremely weak effect for a line seen at $i=30^\circ$, produces some clear effects for $i = 80^\circ$.

%%%%%%%%%%%%%%%%%%%%%%%%%%%%%%%

\begin{table}
 \centering
 \caption{}
\begin{tabular}{|l|lcc|}
\hline
& Parameter & Simulation & Fit \\
\hline
& Energy flux (2-10~keV) & $10^{-9}$~erg/s/cm$^2$ &  \\ 
& Exposure time & 50~ks &  \\
\hline
& $\Gamma$ & 1.6 & free \\
& $q$ & 3 & free \\
& $r_{\rm in}$ & $r_{\rm ISCO}$ & frozen \\
& $r_{\rm out}$ & 400 & frozen \\
& $z$ & 0 & frozen \\
& $\log\xi$ & 3.1 & free \\
& $A_{\rm Fe}$ & 5 & free \\
& $E_{\rm cut}$ & 120~keV & frozen \\
& Reflection fraction & 3 & free \\
\hline
& $\alpha_{13}$ & 0 & free \\
Simulation~1 & $a_*$ & 0.8 & free \\ 
& $i$ & 30$^\circ$ & free \\
\hline
& $\alpha_{13}$ & $-2$ & free \\
Simulation~2 & $a_*$ & 0.8 & free \\ 
& $i$ & 30$^\circ$ & free \\
\hline
& $\alpha_{13}$ & 0 & free \\
Simulation~3 & $a_*$ & 0.8 & free \\ 
& $i$ & 80$^\circ$ & free \\
\hline
\end{tabular}
\vspace{0.3cm}
\tablenotetext{0}{Summary of the values of the parameters employed in our simulations and fits. $\Gamma$ is the photon index of the power-law component, $q$ is the emissivity index, $z$ is the cosmological redshift, $\log\xi$ is the ionization parameter, and $A_{\rm Fe}$ is the iron abundance (in units of Solar iron abundance).} 
\label{tab-sim}
\end{table}

\begin{figure}[t]
\begin{center}
\includegraphics[width=7.5cm]{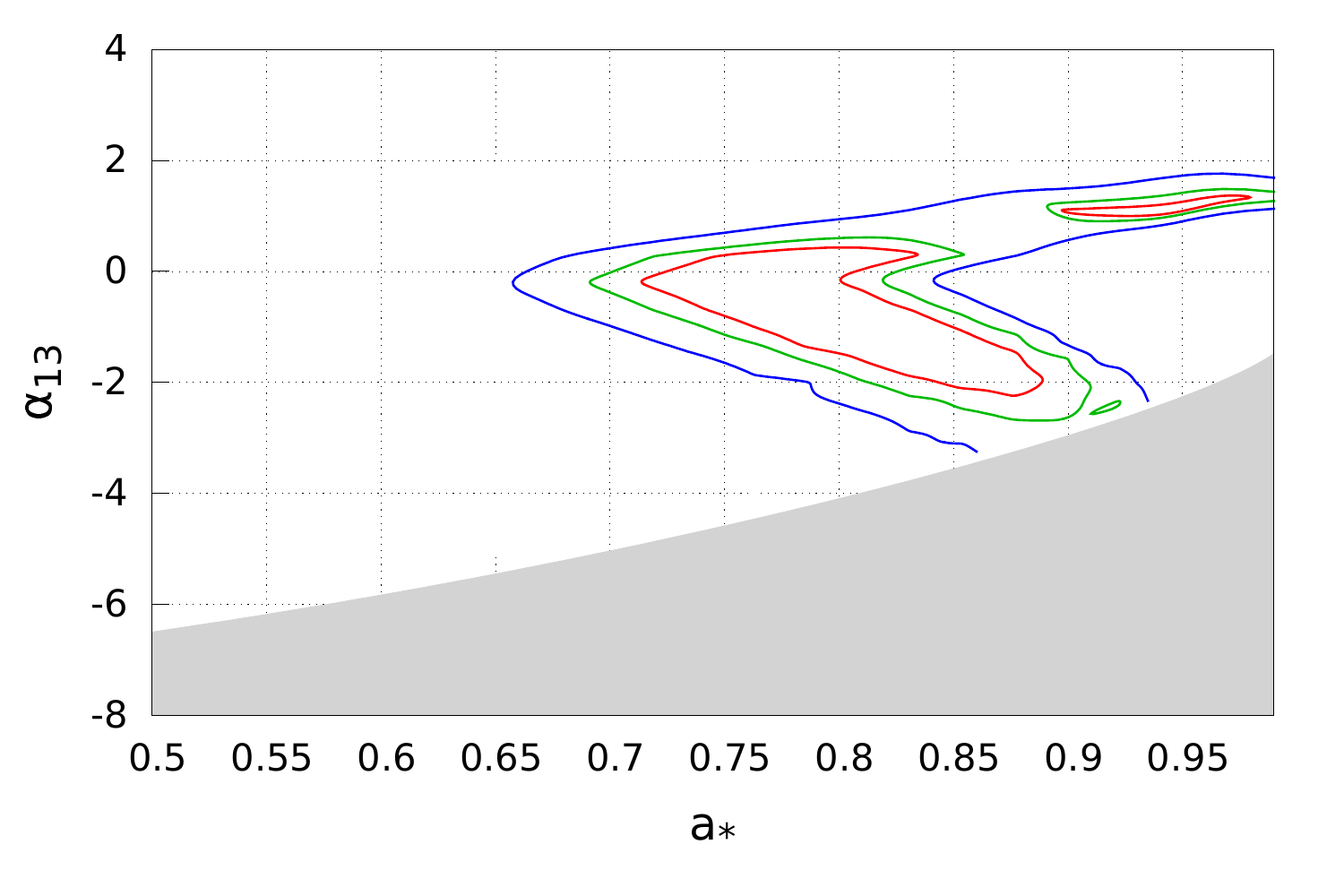}
\hspace{1.0cm}
\includegraphics[width=7.5cm]{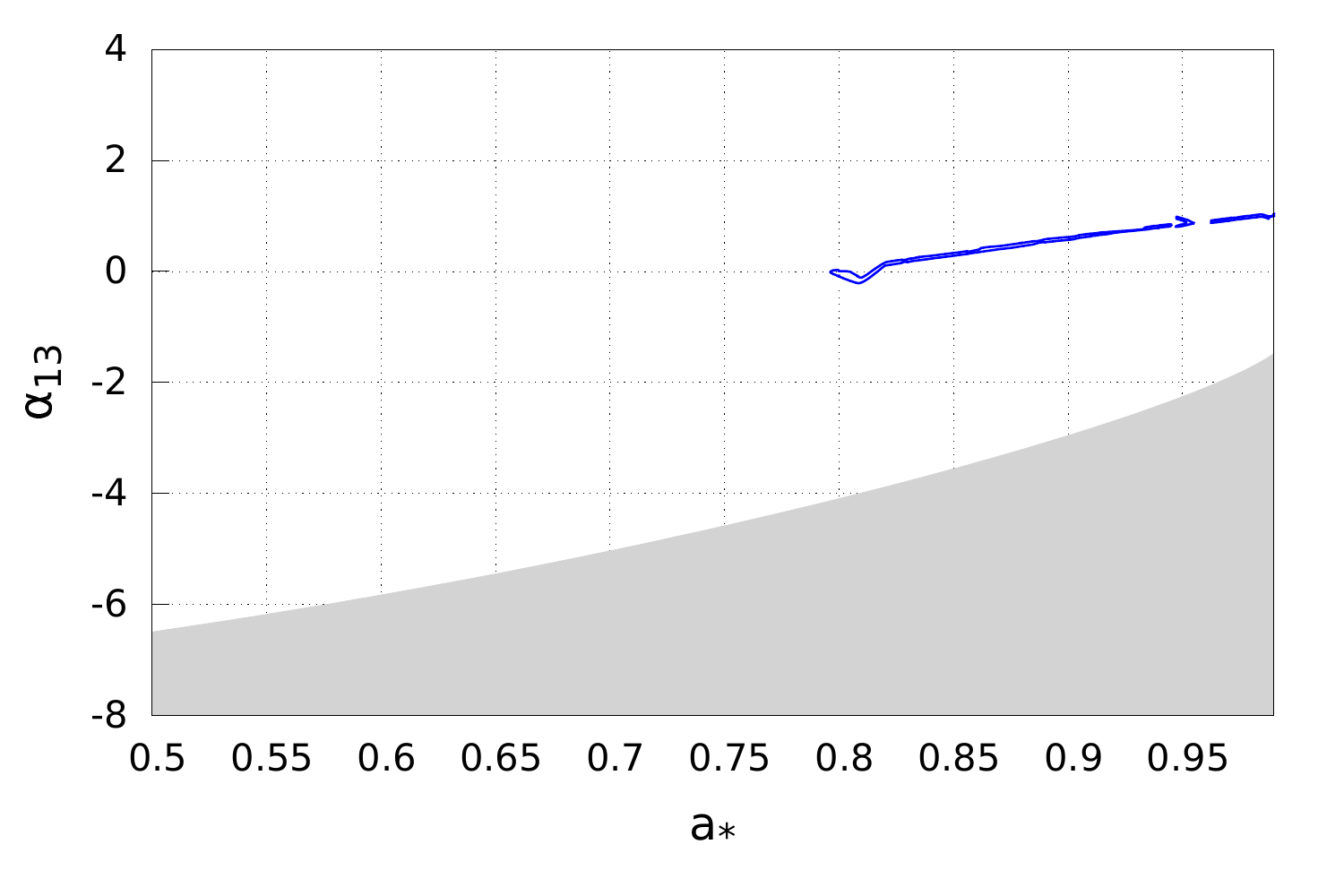}
\end{center}
\caption{1-, 2-, 3-$\sigma$ confidence contours for the spin parameter $a_*$ and the deformation parameter $\alpha_{13}$ from a simulated observation of a bright black hole binary with NuSTAR (left panel) and LAD/eXTP (right panel). The spacetime metric of the simulation has $\alpha_{13} = 0$ (Kerr) and $a_* = 0.8$; the viewing angle is $i = 30^\circ$ (Simulation~1). The grayed region is outside the range prescribed for $\alpha_{13}$ in~(\ref{eq-regularity}) and therefore we restrict our analysis to the regions above. See the text for more details.}
\label{f-relxill1}
\vspace{0.5cm}
\end{figure}

\begin{figure}[t]
\begin{center}
\includegraphics[width=7.5cm]{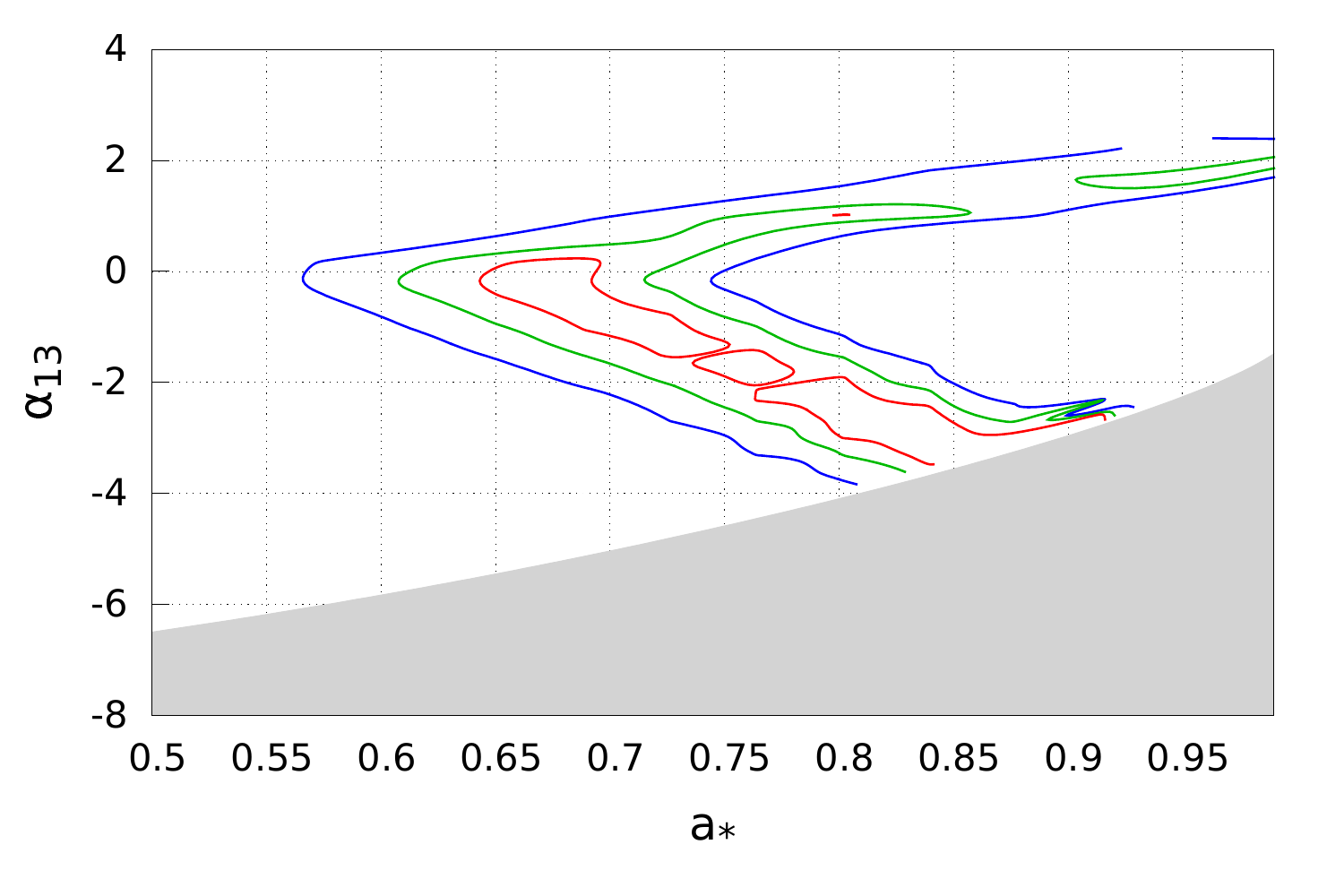}
\hspace{1.0cm}
\includegraphics[width=7.5cm]{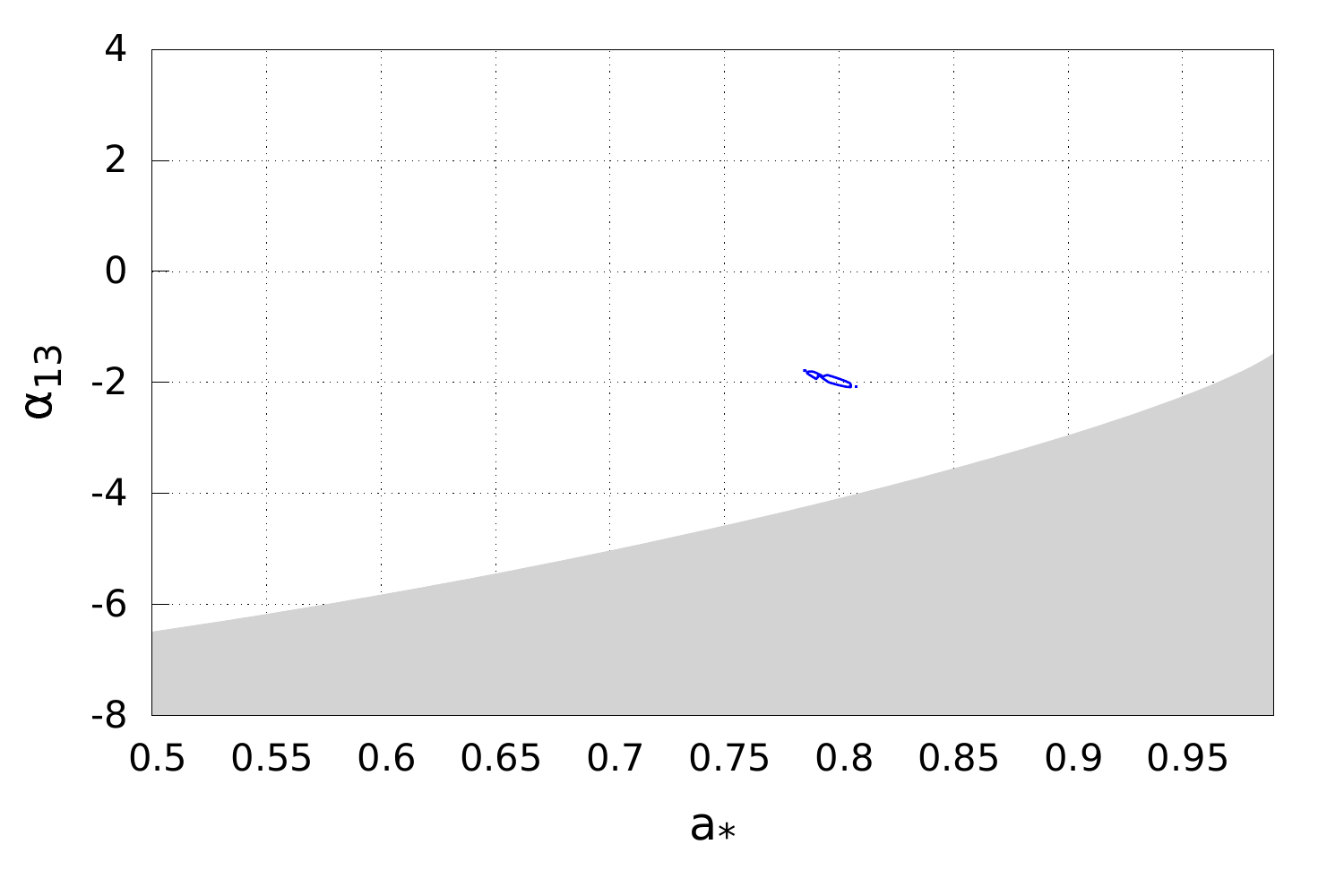}
\end{center}
\caption{As in Fig.~\ref{f-relxill1} for $\alpha_{13} = -2$, $a_* = 0.8$, and $i = 30^\circ$ (Simulation~2). See the text for more details.}
\label{f-relxill2}
\vspace{0.5cm}
\end{figure}
\vspace{0.5cm}
\begin{figure}[t]
\begin{center}
\includegraphics[width=7.5cm]{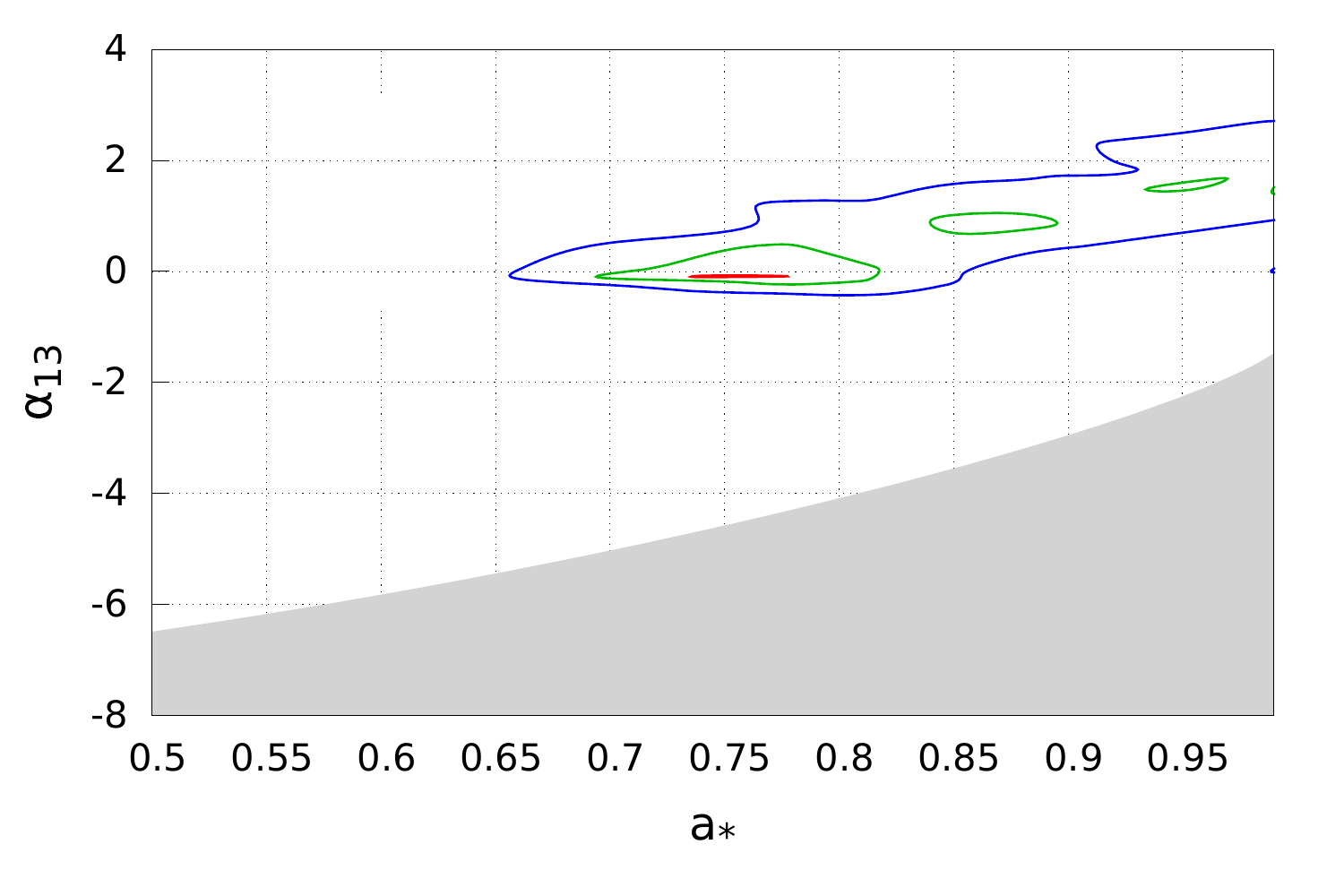}
\hspace{1.0cm}
\includegraphics[width=7.5cm]{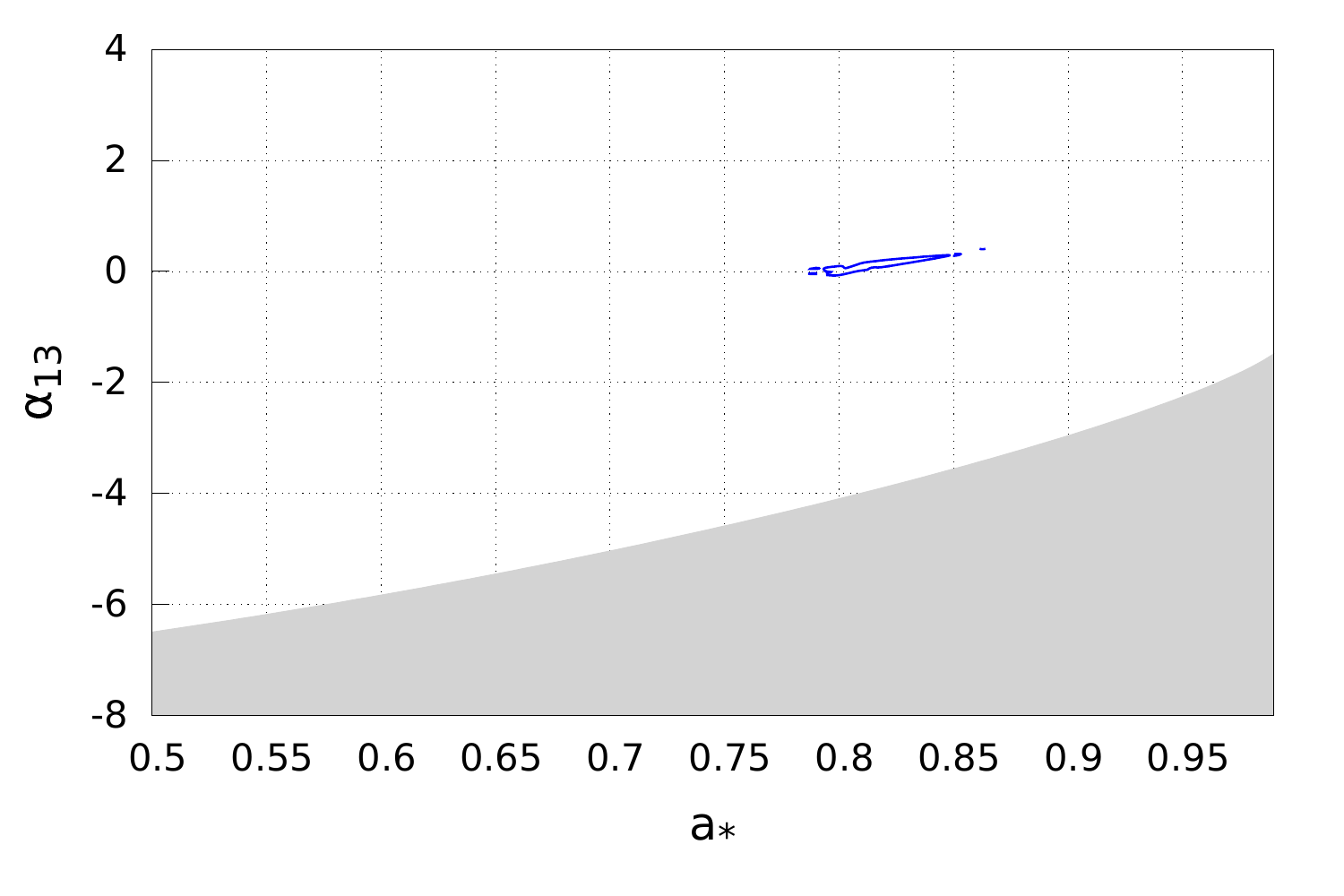}
\end{center}
\caption{As in Fig.~\ref{f-relxill1} for $\alpha_{13} = 0$ (Kerr), $a_* = 0.8$, and $i = 80^\circ$ (Simulation~3). See the text for more details.}
\label{f-relxill3}
\vspace{0.5cm}
\end{figure}

\section{Reflection spectrum}\label{s-new}

In this section, we consider the full reflection spectrum of the accretion disk and we illustrate with some examples how we can constrain the deformation parameters. We simulate observations with NuSTAR and LAD/eXTP to show the constraining power of current and future X-ray missions, respectively. A detailed analysis to study the parameter degeneracy will be presented in a forthcoming paper.

We consider the case of a bright black hole binary, which is expected to be the most suitable source for this kind of test, and we set its energy flux in the 2-10~keV range at $10^{-9}$~erg/s/cm$^2$. The exposure time is 50~ks. The resulting total number of counts is $\sim 10^6$ for NuSTAR and $\sim 10^8$ for LAD/eXTP. We simulate three observations with the extended {\sc relxill}. The values of the input parameters are shown in Tab.~\ref{tab-sim}. The photon index of the continuum is $\Gamma = 1.6$; the spin parameter is always $a_* = 0.8$; the emissivity profile is assumed a simple power-law with index 3, namely $\propto 1/r^3_{\rm e}$; the ionization parameter is $\log\xi = 3.1$ ($\xi$ in units erg~cm/s); the iron abundance is $A_{\rm Fe} = 5$ (in units of Solar iron abundance); the energy cut-off of the continuum is $E_{\rm cut} = 120$~keV; the reflection fraction is chosen to be 3. In Simulation~1, we have a Kerr black hole observed from the viewing angle $i = 30^\circ$. In Simulations~2, we have a non-Kerr black hole with the deformation parameter $\alpha_{13} = -2$ (all the other deformation parameters vanish); the inclination angle is still $i = 30^\circ$. In Simulation~3, we have a Kerr black hole observed from the viewing angle $i = 80^\circ$.

The last column in Tab.~\ref{tab-sim} shows which parameters are free and which are frozen in the fit. Since here we are merely interested in some examples to illustrate the constraints from possible observations with current and future X-ray missions, the initial values of the fit are chosen close to the actual values employed in the simulation. With NuSTAR, we analyze the data in the range 3-70~keV, while in the case of LAD/eXTP the range is 1-70~keV. Fig.~\ref{f-relxill1} shows the map of $\Delta\chi^2$ for the spin parameter and the deformation parameter $\alpha_{13}$ for Simulation~1; the left panel is the result from the simulation with NuSTAR and the right panel is that for LAD/eXTP. The red, green, and blue curves indicate, respectively, the 1-, 2-, 3-$\sigma$ limits. For the simulation with LAD/eXTP, we only show the blue 3-$\sigma$ contour because the allowed region is extremely thin. The gray region is not analyzed because the spacetimes there do not meet the condition on $\alpha_{13}$ in~(\ref{eq-regularity}). The confidence contours for Simulation~2 are shown in Fig.~\ref{f-relxill2} and those for Simulation~3 in Fig.~\ref{f-relxill3}.

The degeneracy between the spin and the deformation parameter $\alpha_{13}$ is clear. While this depends on the choice of the deformation parameter, so in our case $\alpha_{13}$, it is quite a common feature, especially when the deformation parameter has a strong impact on the value of the ISCO radius. In Fig.~\ref{f-relxill1}, the contours for NuSTAR show that we could potentially find a large spin for either positive or negative deformations. The negative branch is removed with LAD/eXTP, but despite of the very small uncertainty, the positive branch is still there, which means that for small inclinations the problem of degeneracy may persist despite the large effective area. In Fig.~\ref{f-relxill3}, the inclination angle is large, which maximizes the relativistic effects and helps to break the parameter degeneracy. While this looks indeed the best case for NuSTAR, the problem of degeneracy persists.

The remarkable difference between the constraining power of NuSTAR and LAD/eXTP was already pointed out in~\citet{yy16}. We note that LAD/eXTP can potentially provide so stringent constraints on the deformation parameters that the choice of the correct theoretical model, e.g. the choice of the form of the emissivity profile, will be crucial to get reliable constraints on the spacetime metric.

%%%%%%%%%%%%%%%%%%%%%%%%%%%%%%%

\section{Summary and conclusions}\label{s-7}

In this Paper, we present the first X-ray reflection model for testing the spacetime metric around astrophysical black holes. Previous work suggests that the reflection method is quite a promising technique to test the Kerr black hole hypothesis with electromagnetic radiation. However, current studies employ simplified models. In the best cases, the X-ray spectrum is approximated by a power law with an iron line. Similar models can work for preliminary studies, but they are definitively inadequate to perform precise tests of general relativity in the strong gravity regime.

{\sc relxill} is currently the most sophisticated model to fit the X-ray reflection spectrum of black holes under the assumption that the spacetime is described by the Kerr solution. It results from the combination of the relativistic convolution model for the Kerr metric {\sc relconv} and the reflection code for the local spectrum {\sc xillver}. By calculating the transfer function for a generic background, we have a new relativistic convolution model to replace {\sc relconv}. After merging our new relativistic convolution model with {\sc xillver}, we obtain the extension of {\sc relxill} to generic stationary, axisymmetric, and asymptotically flat black hole spacetime.

We have described our new code and the relevant formulas for the calculation of the transfer function. We have shown that our calculations reach the necessary accuracy for our tests. We have simulated some observations of a bright black hole binary with NuSTAR and LAD/eXTP to illustrate the constraining power of current and future X-ray missions. The current version of the code adopts the Johannsen metric, but it is straightforward to employ any other stationary, axisymmetric, and asymptotically flat black hole metric. Work on other non-Kerr metrics, such as that proposed in~\citet{ref-krz}, is currently underway. In a forthcoming paper, we will apply our new model to a specific source for constraining the deformation parameters of the Johannsen metric from available X-ray data.

%%%%%%%%%%%%%%%%%%%%%%%%%%%%%%%

\begin{acknowledgments}
We thank Jiachen Jiang for useful discussions and suggestions. C.B. and S.N. were supported by the NSFC (grants U1531117 and 11305038), Fudan University (Grant No.~IDH1512060), and the Thousand Young Talents Program. C.B. also acknowledges the support from the Alexander von Humboldt Foundation. A.C.-A. acknowledges funding from the Fundaci\'on Universitaria Konrad Lorenz (Project 5INV1) and thanks the Department of Physics at Fudan University for hospitality during his visit. J.A.G. acknowledges the support of a CGPS grant from the Smithsonian Institution.
\end{acknowledgments}

%%%%%%%%%%%%%%%%%%%%%%%%%%%%%%%

\appendix

\section{Calculations of the transfer function}\label{s-3}

In order to calculate the transfer function in Eq.~(\ref{eq-trf}), we have to map the emission points in the disk onto the image plane of the distant observer. This can be achieved by calculating the photon trajectories connecting the emission points to the detection points.

\subsection{Photon initial conditions}

\begin{figure}[t]
\begin{center}
\includegraphics[angle=270,width=10cm]{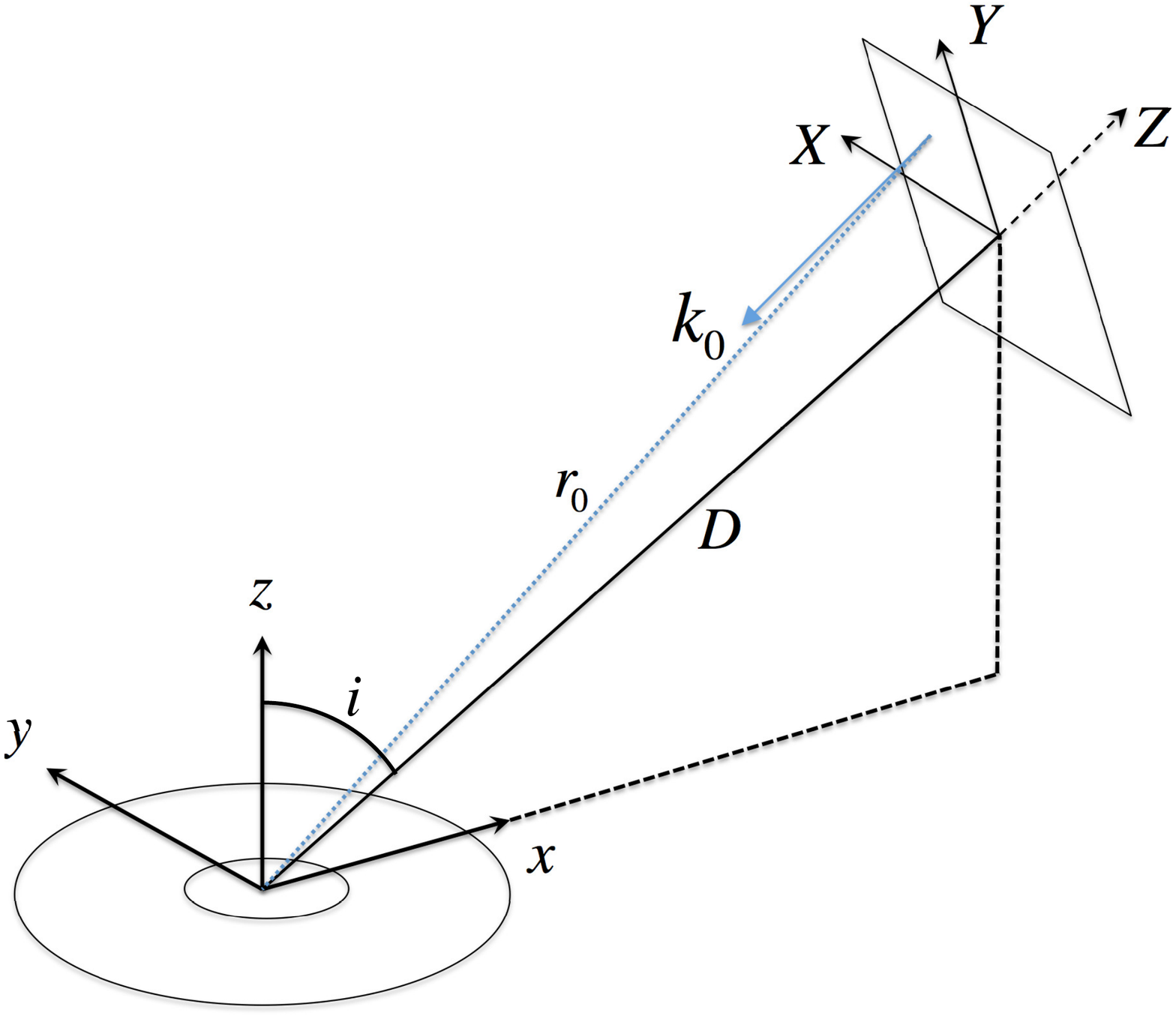}
\end{center}
\caption{The Cartesian coordinates $(x,y,z)$ are centred at the black hole, while the Cartesian coordinates $(X,Y,Z)$ are for the image plane of the distant observer, who is located at the distant $D$ from the black hole and with an inclination angle $i$.}
\label{f-setup}
\vspace{0.5cm}
\end{figure}

The first step is to write the photon initial conditions in the image plane of the distant observer~\citep{jp-image,code-cfm}. Let us consider a black hole surrounded by an accretion disk and an observer at the distant $D$ from the black hole and with the viewing angle $i$, as sketched in Fig.~\ref{f-setup}. The image plane of the distant observer is provided with a system of Cartesian coordinates $(X,Y,Z)$. Another system of Cartesian coordinates $(x,y,z)$ is centred at the black hole. The two Cartesian coordinates are related by 
\be
x &=& D \sin i - Y \cos i + Z \sin i \, , \nonumber\\
y &=& X \, , \nonumber\\
z &=& D \cos i + Y \sin i + Z \cos i \, .
\ee

Let us assume the black hole metric is expressed in spherical-like coordinates. Far from the compact object, the spatial coordinates reduce to the usual spherical coordinates in flat spacetime and they are related to $(x,y,z)$ by
\be
r &=& \sqrt{x^2 + y^2 + z^2} \, , \nonumber\\ 
\theta &=& \arccos \left(\frac{z}{r}\right) \, , \nonumber\\
\phi &=& \arctan \left(\frac{y}{x}\right) \, .
\ee

Let us consider a photon at the position $(X_0, Y_0, 0)$ and with 3-momentum ${\bf k}_0 = - k_0 \hat{Z}$ perpendicular to the image plane. The initial conditions for the photon position are
\be\label{eq-m-initial-x}
t_0 &=& 0 \, , \nonumber\\
r_0 &=& \sqrt{X_0^2 + Y_0^2 + D^2} \, , \nonumber\\
\theta_0 &=& \arccos \frac{Y_0 \sin i + D \cos i}{r_0} \, , \nonumber\\
\phi_0 &=& \arctan \frac{X_0}{D \sin i - Y_0 \cos i} \, .
\ee
The photon 4-momentum is $k^\mu = (\partial x^\mu/\partial \tilde{x}^\alpha) \tilde{k}^\alpha$, where $\tilde{k}^\alpha = (k_0,0,0,-k_0)$ is the photon 4-momentum in the Cartesian coordinates, and we find
\be\label{eq-m-initial-k}
k^r_0 &=& - \frac{D}{r_0} |{\bf k}_0| \, , \nonumber\\
k^\theta_0 &=& \frac{\cos i - \left(Y_0 \sin i + D \cos i\right) 
\frac{D}{r_0^2}}{\sqrt{X_0^2 + (D \sin i - Y_0 \cos i)^2}} |{\bf k}_0| \, , \nonumber\\
k^\phi_0 &=& \frac{X_0 \sin i}{X_0^2 + (D \sin i - Y_0 \cos i)^2} |{\bf k}_0| \, .
\ee
$k^t_0$ can be obtained from the condition $g_{\mu\nu}k^\mu k^\nu = 0$ with the metric tensor of a flat spacetime, so 
\be\label{eq-m-initial-kbis}
k^t_0 = \sqrt{\left(k^r_0\right)^2 + r^2_0  \left(k^\theta_0\right)^2 + r_0^2 \sin^2\theta_0  (k^\phi_0)^2} \, .
\ee

\subsection{Photon trajectory}

With the photon initial conditions~(\ref{eq-m-initial-x}), (\ref{eq-m-initial-k}), and (\ref{eq-m-initial-kbis}), we can integrate the geodesic equations backward in time from any detection point $(X_0, Y_0, 0)$ in the image plane of the distant observer to the emission point in the disk:
\be
\frac{d^2 x^\mu}{d\tau^2} + \Gamma^\mu_{\nu\rho} 
\frac{d x^\nu}{d\tau}
\frac{d x^\rho}{d\tau} = 0 \, ,
\ee
where $\tau$ is an affine parameter. In the case of the Kerr metric, it is not necessary to directly integrate the geodesic equations and the calculations are thus different~\citep{cun75,spe95}. In the Kerr metric in Boyer-Lindquist coordinates, the equations of motion are separable and we can restrict the attention to the motion in the $(r,\theta)$ plane; the corresponding equations can be solved in terms of elliptic integrals.

\subsection{Accretion disk}

The integration of the geodesic equations stops when the photon either hits the accretion disk or misses it. In the latter case, the photon either hits the black hole or crosses the equatorial plane between the black hole and the inner edge of the disk. In our model, only the primary image of the accretion disk is take into account, and there is not emission between the black hole and the inner edge of the disk. The inner edge of the disk is assumed at the ISCO radius and, for a generic stationary, axisymmetric, and asymptotically flat spacetime can be inferred as follows. We write the line element in the canonical form, namely 
\be
ds^2 = g_{tt} dt^2 + 2 g_{t\phi} dt d\phi + g_{rr} dr^2 
+ g_{\theta\theta} d\theta^2 + g_{\phi\phi} d\phi^2 \, ,
\ee
where the metric coefficients are independent of $t$ and $\phi$. The motion of a test-particle in the metric background is governed by the Lagrangian
\be
{\mathcal L} = \frac{1}{2} g_{\mu\nu} \dot{x}^\mu \dot{x}^\nu \, ,
\ee
where $\dot{} = d/d\tau$. Since the metric is independent of the coordinates $t$ and $\phi$, we have two constants of motion, namely the specific energy at infinity $E$ and the axial component of the specific angular momentum at infinity $L_z$:
\be\label{eq-th-l000a}
\frac{d}{d\tau} \frac{\partial {\mathcal L}}{\partial \dot{t}} 
- \frac{\partial {\mathcal L}}{\partial t} = 0 
&\Rightarrow& p_t \equiv \frac{\partial {\mathcal L}}{\partial \dot{t}}
= g_{tt} \dot{t} + g_{t\phi} \dot{\phi}= - E \, , \\
\frac{d}{d\tau} \frac{\partial {\mathcal L}}{\partial \dot{\phi}} 
- \frac{\partial {\mathcal L}}{\partial \phi} = 0 
&\Rightarrow& p_\phi \equiv \frac{\partial {\mathcal L}}{\partial \dot{\phi}}
= g_{t\phi} \dot{t} + g_{\phi\phi} \dot{\phi} = L_z \, .
\label{eq-th-l000}
\ee
The term ``specific'' is used to indicate that $E$ and $L_z$ are, respectively, the energy and angular momentum per unit rest-mass. Eqs.~(\ref{eq-th-l000a}) and (\ref{eq-th-l000}) can be solved to find the $t$- and the $\phi$-component of the 4-velocity of the test-particle
\be\label{eq-tdot-phidot}
\dot{t} = \frac{E g_{\phi\phi} + L_z g_{t\phi}}{g^2_{t\phi} - g_{tt} g_{\phi\phi}} \, , \qquad
\dot{\phi} = - \frac{E g_{t\phi} + L_z g_{tt}}{g^2_{t\phi} - g_{tt} g_{\phi\phi}} \, .
\ee

The accretion disk is described by the Novikov-Thorne model~\citep{nt1,nt2}. The disk is in the equatorial plane perpendicular to the black hole spin. The particles of the gas follow nearly geodesic, equatorial, circular orbits. We write the geodesic equations as
\be\label{eq-circular}
\frac{d}{d\tau} \left( g_{\mu\nu} \dot{x}^\nu \right) = 
\frac{1}{2} \left( \partial_\mu g_{\nu\rho} \right) \dot{x}^\nu \dot{x}^\rho \, . 
\ee
Since $\dot{r} = \dot{\theta} = \ddot{r} = 0$ for equatorial circular orbits, the radial component of Eq.~(\ref{eq-circular}) reduces to
\be
\left(\partial_r g_{tt}\right) \dot{t}^2 
+ 2 \left(\partial_r g_{t\phi}\right) \dot{t} \dot{\phi} 
+ \left(\partial_r g_{\phi\phi}\right) \dot{\phi}^2 = 0 \, .
\ee
The angular velocity $\Omega = \dot{\phi}/\dot{t}$ is
\be\label{eq-omega-g}
\Omega = \frac{- \partial_r g_{t\phi} 
\pm \sqrt{\left(\partial_r g_{t\phi}\right)^2 - 
\left(\partial_r g_{tt}\right) \left(\partial_r g_{\phi\phi}\right)}}{\partial_r g_{\phi\phi}} \, ,
\ee
where the upper (lower) sign refers to corotating (counterrotating) orbits, namely orbits with angular momentum parallel (antiparallel) to the spin of the central object.

From $g_{\mu\nu} \dot{x}^\mu \dot{x}^\nu = -1$ with $\dot{r} = \dot{\theta} = 0$, we can write
\be\label{eq-ttt}
\dot{t} = \frac{1}{\sqrt{- g_{tt} - 2 \Omega g_{t\phi} - \Omega^2 g_{\phi\phi}}} \, .
\ee
Eq.~(\ref{eq-th-l000a}) becomes
\be\label{eq-th-e}
E &=& - \left( g_{tt} + \Omega g_{t\phi} \right) \dot{t} \nonumber\\
&=& - \frac{g_{tt} + \Omega g_{t\phi}}{\sqrt{- g_{tt} 
- 2 \Omega g_{t\phi} - \Omega^2 g_{\phi\phi}}}\, .
\ee
In the same way, Eq.~(\ref{eq-th-l000}) becomes 
\be\label{eq-th-l}
L_z &=& \left( g_{t\phi} + \Omega g_{\phi\phi} \right) \dot{t} \nonumber\\
&=& \frac{g_{t\phi} + \Omega g_{\phi\phi}}{\sqrt{- g_{tt} 
- 2 \Omega g_{t\phi} - \Omega^2 g_{\phi\phi}}} \, .
\ee

From $g_{\mu\nu} \dot{x}^\mu \dot{x}^\nu = -1$ and with the use of Eq.~(\ref{eq-tdot-phidot}), we can now write
\be\label{eq-veff}
g_{rr} \dot{r}^2 + g_{\theta\theta}^2 \dot{\theta}^2 = V_{\rm eff} (r, \theta, E, L_z) \, ,
\ee
where $V_{\rm eff} (r, \theta, E, L_z)$ is the effective potential of the test-particle with energy $E$ and axial component of the angular momentum $L_z$
\be\label{eq-veff-def}
V_{\rm eff} = \frac{E^2 g_{\phi\phi} + 2 E L_z g_{t\phi} 
+ L^2_z g_{tt}}{g^2_{t\phi} - g_{tt} g_{\phi\phi}} - 1 \, .
\ee

In the case of equatorial circular orbits, $\dot{r} = \dot{\theta} = \ddot{r} = \ddot{\theta} = 0$ and therefore $V_{\rm eff} = \partial_r V_{\rm eff} = \partial_\theta V_{\rm eff} = 0$. The orbit is radially (vertically) stable if $\partial^2_r V_{\rm eff} < 0$ ($\partial^2_\theta V_{\rm eff} < 0$) and radially (vertically) unstable if $\partial^2_r V_{\rm eff} > 0$ ($\partial^2_\theta V_{\rm eff} > 0$). The ISCO radius $r_{\rm ISCO}$ is given by
\be
\partial^2_r V_{\rm eff} = 0 \;\; {\rm or} \;\; \partial^2_\theta V_{\rm eff} = 0
\quad \Rightarrow \quad r = r_{\rm ISCO} \, .
\ee

\subsection{Redshift factor and emission angle}

Once we know the emission point in the disk, we can evaluate the redshift factor $g$ and the emission angle $\vartheta_{\rm e}$. The redshift factor $g$ is
\be\label{eq-gg}
g = \frac{\nu_{\rm o}}{\nu_{\rm e}}
= \frac{-u^\mu_{\rm o} k_\mu}{-u^\nu_{\rm e} k_\nu} \, ,
\ee
where $u^\mu_{\rm o} = (1,0,0,0)$ is the 4-velocity of the distant observer, $k^\mu$ is the 4-momentum of the photon, and $u^\nu_{\rm e} = u^t_{\rm e} (1,0,0,\Omega)$ is the 4-velocity of the particles of the gas. $u^t_{\rm e} = \dot{t}$, which is given by Eq.~(\ref{eq-ttt}). Plugging Eq.~(\ref{eq-ttt}) into Eq.~(\ref{eq-gg}), we obtain
\be\label{eq-utgg}
g = \frac{\sqrt{-g_{tt} - 2g_{t\phi}\Omega - 
g_{\phi\phi}\Omega^2}}{1 - \lambda \Omega} \, ,
\ee
where $\lambda = - k_\phi/k_t$ is a constant of motion along the photon trajectory and can be evaluated from the initial conditions.

If the local spectrum depends on the emission angle $\vartheta_{\rm e}$, the latter must be rewritten in terms of the emission radius and redshift factor. The normal of the disk is
\be
n^\mu = \Big( 0, 0, \sqrt{g^{\theta\theta}} , 0 \Big) \Big|_{r_{\rm e}, \theta_{\rm e} = \pi/2} \, , 
\ee
and therefore the cosine of the emission angle $\vartheta_{\rm e}$ is
\be
\cos \vartheta_{\rm e} = \frac{n^\mu k_\mu}{u^\nu_{\rm e} k_\nu} \Big|_{\rm e}
= \sqrt{g^{\theta\theta}} 
\frac{\sqrt{ - g_{tt} - 2 g_{t\phi} \Omega - g_{\phi\phi} \Omega^2}}{1 - \lambda \Omega}
\frac{k_\theta}{k_t} \, ,
\ee
where $k_\theta$ is the $\theta$-component of the 4-momentum of the photon at the point of emission in the disk and, in the general case, it is determined at the end of the geodesic integration.

At the end of the integration of the photon trajectory we have $r_{\rm e} = r_{\rm e}(X,Y)$, $g = g(X,Y)$, and $\vartheta_{\rm e} = \vartheta_{\rm e}(X,Y)$. From the first two relations, it is possible to numerically compute the Jacobian in the transfer function
\be
\left| \frac{\partial(X,Y)}{\partial(g^*,r_{\rm e})} \right| 
= \left(g_{\rm max} - g_{\rm min}\right)
\left| \frac{\partial X}{\partial g} \frac{\partial Y}{\partial r_{\rm e}} 
- \frac{\partial X}{\partial r_{\rm e}} \frac{\partial Y}{\partial g}\right| \, .
\ee
This completes the calculations of the transfer function $f$ for a specific background metric. If we know the local spectrum of the radiation $I_{\rm e}$, we can obtain the observed flux via Eq.~(\ref{eq-Fobs}).

%%%%%%%%%%%%%%%%%%%%%%%%%%%%%%%

\end{document}